\documentclass[a4paper,fleqn,usenatbib]{mnras}

\usepackage{newtxtext,newtxmath}
\usepackage[flushleft]{threeparttable}
\usepackage{times,epsfig,natbib,amssymb,amsmath,graphics,longtable,lscape,threeparttable}

\usepackage[T1]{fontenc}
\usepackage{ae,aecompl}

\usepackage{graphicx}	
\usepackage{amsmath}	
\usepackage{nccmath}
\usepackage{amssymb}	
\usepackage{pdflscape}	
\usepackage{multicol}
\usepackage{threeparttablex}

\title[Berkeley Type II supernova sample]{The Berkeley sample of Type II supernovae: $BVRI$ light curves and spectroscopy of 55 SNe~II}

\author[de Jaeger et al.]
{T. de Jaeger$^{1,2}$\thanks{E-mail: tdejaeger@berkeley.edu},
W. Zheng$^{1}$,
B. E. Stahl$^{1,3,4}$,
A. V. Filippenko$^{1,5}$, 
T. G. Brink$^{1}$,
\newauthor
A. Bigley$^{1}$,
K. Blanchard$^{1}$,
P. K. Blanchard$^{6}$,
J. Bradley$^{1}$,
S. K. Cargill$^{1}$,
C. Casper$^{1}$,
\newauthor
S. B. Cenko$^{7,8}$,
S. Channa$^{1}$,
B. Y. Choi$^{1}$,
K. I. Clubb$^{1}$,
B. E. Cobb$^{9}$,
D. Cohen$^{10}$,
\newauthor
M. de Kouchkovsky$^{1}$,
M. Ellison$^{1}$,
E. Falcon$^{1}$,
O. D. Fox$^{11}$,
K. Fuller$^{1}$,
\newauthor
M. Ganeshalingam$^{12}$,
C. Gould$^{1}$,
M. L. Graham$^{13}$,
G. Halevi$^{1,14}$,
K. T. Hayakawa$^{10}$,
\newauthor 
J. Hestenes$^{1}$,
M. P. Hyland$^{1}$,
B. Jeffers$^{1}$,
N. Joubert$^{15,1}$,
M. T. Kandrashoff$^{1}$,
\newauthor
P. L. Kelly$^{16,1}$,
H. Kim$^{1}$,
M. Kim$^{1}$,
S. Kumar$^{17,1}$,
E. J. Leonard$^{18}$,
G. Z. Li$^{19}$,
\newauthor
T. B. Lowe$^{20,21}$,
P. Lu$^{1,10}$,
M. Mason$^{1,22}$,
K. J. McAllister$^{1,4}$,
J. C. Mauerhan$^{23,1}$,
\newauthor
M. Modjaz$^{24}$,
J. Molloy$^{1}$,
D. A. Perley$^{25}$,
K. Pina$^{1}$,
D. Poznanski$^{26}$,
T. W. Ross$^{1}$,
\newauthor
I. Shivvers$^{1}$,
J. M. Silverman$^{27}$,
C. Soler$^{1}$,
S. Stegman$^{1}$,
S. Taylor$^{1}$,
K. Tang$^{1}$,
A. Wilkins$^{1}$,
\newauthor
Xiaofeng Wang$^{28}$,
Xianggao Wang$^{29}$,
H. Yuk$^{30}$,
S. Yunus$^{1}$,
K. D. Zhang$^{1}$
\\
\small
\\
$^{1}$Department of Astronomy, University of California, Berkeley, CA 94720-3411, USA.\\
$^{2}$Bengier Postdoctoral Fellow.\\
$^{3}$Marc J. Staley Graduate Fellow.\\
$^{4}$Department of Physics, University of California, Berkeley, CA 94720-7300, USA.\\
$^{5}$Miller Senior Fellow, Miller Institute for Basic Research in Science, University of California, Berkeley, CA 94720, USA.\\
$^{6}$Harvard-Smithsonian Center for Astrophysics, 60 Garden Street, Cambridge, MA 02138, USA.\\
$^{7}$Astrophysics Science Division, NASA Goddard Space Flight Center, MC 661, Greenbelt, MD 20771, USA.\\
$^{8}$Joint Space-Science Institute, University of Maryland, College Park, MD 20742, USA.\\
$^{9}$Department of Physics, The George Washington University, Washington, DC 20052, USA.\\
$^{10}$Department of Physics and Astronomy, University of California, Los Angeles, CA 90095, USA.\\
$^{11}$Space Telescope Science Institute, 3700 San Martin Drive, Baltimore, MD 21218, USA.\\ 
$^{12}$Energy Analysis and Environmental Impacts Division, Lawrence Berkeley National Laboratory, 1 Cyclotron Road, Berkeley, CA 94720, USA.\\
$^{13}$University of Washington, Department of Astronomy Box 351580 Seattle WA 98195-1580, USA.\\
$^{14}$Department of Astrophysical Sciences, Princeton University, Princeton, NJ 08542, USA.\\ 
$^{15}$Gates Computer Science Building 353 Serra Mall Stanford, CA 94305, USA.\\              
$^{16}$School of Physics and Astronomy, University of Minnesota, 116 Church Street SE, Minneapolis,MN 55455, USA.\\
$^{17}$Department of Physics, Florida State University, Tallahassee, FL 32306, USA.\\
$^{18}$Department of Earth, Planetary, and Space Sciences at the University of California, Los Angeles, CA 90095, USA.\\
$^{19}$Department of Mechanical and Aerospace Engineering, University of California, Los Angeles, CA 90095, USA.\\
$^{20}$Institute for Astronomy, University of Hawaii, 2680 Woodlawn Drive, Honolulu, HI 96822, USA.\\
$^{21}$Lick Observatory, P.O. Box 85, Mount Hamilton, CA 95140, USA.\\
$^{22}$Department of Physics and Astronomy, University of Wyoming, 1000 E. University Ave. Laramie, WY 82071, USA.\\
$^{23}$The Aerospace Corporation, 2310 E. El Segundo Blvd., El Segundo, CA 90245, USA.\\
$^{24}$Center for Cosmology and Particle Physics, New York University, 4 Washington Place, New York, NY 10003, USA.\\
$^{25}$Astrophysics Research Institute, Liverpool John Moores University, IC2, Liverpool Science Park, 146 Brownlow Hill, Liverpool L3 5RF, UK.\\
$^{26}$School of Physics and Astronomy, Tel-Aviv University, Tel Aviv 69978, Israel.\\ 
$^{27}$Samba TV, San Francisco, CA 94107, USA.\\ 
$^{28}$Physics Department and Astronomy Department, Tsinghua University, Beijing ,100084, China.\\ 
$^{29}$Department of Physics, Guangxi University, Nanning 530004, China.\\    
$^{30}$Department of Physics and Astronomy, University of Oklahoma, 440 W Brooks St, Norman, OK 73019, USA.\\
\\    
}

\date{}

\pubyear{2019}

\begin{document}
\label{firstpage}
\pagerange{\pageref{firstpage}--\pageref{lastpage}}

\maketitle

\newpage

\begin{abstract}

In this work, $BVRI$ light curves of 55 Type II supernovae (SNe~II) from the Lick Observatory Supernova Search program obtained with the Katzman Automatic Imaging Telescope and the 1~m Nickel telescope from 2006 to 2018 are presented. Additionally, more than 150 spectra gathered with the 3~m Shane telescope are published. We conduct an analyse of the peak absolute magnitudes, decline rates, and time durations of different phases of the light and colour curves. Typically, our light curves are sampled with a median cadence of 5.5 days for a total of 5093 photometric points. In average $V$-band plateau declines with a rate of 1.29 mag (100 days)$^{-1}$, which is consistent with previously published samples. For each band, the plateau slope correlates with the plateau length and the absolute peak magnitude: SNe~II with steeper decline have shorter plateau duration and are brighter. A time-evolution analysis of spectral lines in term of velocities and pseudoequivalent widths is also presented in this paper. Our spectroscopic sample ranges between 1 and 200 days post-explosion and has a median ejecta expansion velocity at 50 days post-explosion of 6500 km s$^{-1}$ (H$\alpha$ line) and a standard dispersion of 2000 km s$^{-1}$. Nebular spectra are in good agreement with theoretical models using a progenitor star having a mass $<16$ ${\rm M}_{\odot}$. All the data are available to the community and will help to understand SN~II diversity better, and therefore to improve their utility as cosmological distance indicators.
\end{abstract}



\section{Introduction}

Type I and Type II supernova (SN) classification was initially established by \citet{min41} on the presence or absence of Balmer features in their spectra (see \citealt{filippenko97} for a review). Type II supernovae (hereafter SNe~II) are known to be the final explosion of a massive star with an extensive hydrogen envelope (\citealt{smartt15} for a review). 

The majority of the SN~II progenitors have been constrained first using hydrodynamical models \citep{grassberg71,falk77} and local host-galaxy environment studies \citep{huang1987,vandyk92}, and then later confirmed by direct progenitor detections \citep{vandyk03,smartt09a,fraser12,smartt15,vandyk19}. It is now well accepted that SN~II progenitors are the explosion of only one stellar population (red supergiants) with a zero-age main sequence mass between 8 ${\rm M}_{\odot}$ and $\sim$ 20 ${\rm M}_{\odot}$.

Based on photometric properties, SNe~II were classified into two subgroups: SNe~IIP characterised by a phase of constant luminosity and SNe~IIL with a linear light-curve decline \citep{barbon79}. However, recently, large SN~II sample studies have questioned this sub-classification and have suggested that the SN~II family forms only one continuous group \citep{anderson14a,sanders15,valenti16,galbany16a,dejaeger18a}. Therefore, in this manuscript, SNe~IIP and SNe~II are referred to as SNe~II.

As SN~II progenitors are better understood than any other type of SN (e.g., no direct for SN~Ia progenitor) and because SNe~II are the most abundant SN type in nature ($\sim60$\% \citealt{li2011}), over the last two decades the SN community has demonstrated a growing interest in studying their properties and using them as metallicity \citep{dessart14,anderson16a} or standard candles (e.g., \citealt{hamuy02}). SN~II standardisation using different methods, has shown promising results to measure extragalactic distances: the ``expanding photosphere method'' \citep{kirshner74}, the ``standard candle method'' \citep{hamuy02,nugent06,poznanski09,olivares10,andrea10,poznanski10,dejaeger17b,gall17}, the ``photospheric magnitude method'' \citep{rodriguez14,rodriguez19a}, and the ``photometric colour method'' \citep{dejaeger15b,dejaeger17a}. Moreover, techniques for measuring extragalactic distances using independent methods (such as those afforded by SNe~II) have grown increasingly important in light of recent results showing 4.4$\sigma$ disagreement between local measurements (using SNe~Ia; \citealt{riess16,riess18a,riess19}) of the local Hubble-Lemaitre constant and that inferred from the cosmic microwave background radiation (CMBR) assuming a $\Lambda$CDM cosmology \citep{planck16}.

However, the distance precision derived using SNe~II is still worse than that obtained with SNe~Ia. The dispersion could arise from intrinsic progenitor properties like the mass of the H envelope, the metallicity, the radius, or the characteristics of circumstellar material (CSM) around the progenitor. For example, in the last few years, several studies have shown that the majority of SNe~II at early epochs present evidence of CSM interactions \citep[e.g.,][]{khazov16,morozova16,yaron17,moriya17,dessart17,morozova17,forster18}. Moreover, some SNe~II with strong CSM interaction have proven to be poor standard candles \citep{dejaeger18a}.

Even if individual SN~II studies can be found in the literature as for example: SN~1999em \citep{hamuy01,leonard02,elmhamdi03}, SN~1999gi \citep{leonard02b}, SN~2004et \citep{sahu06,maguire10b}, SN~2005cs \citep{pastorello09}, SN~2013by \citep{valenti15}, SN~2013ej \citep{valenti14,bose15,huang15,mauerhan16,dhungana16}), and SN~2016esw \citep{dejaeger18b}, only a small fraction of large SN~II samples have been published \citep{hamuy03a,arcavi13,anderson14a,faran14a,faran14b,spiro14a,sanders15,galbany16a,valenti16,hicken17,gutierrez17b}. Investigating large samples is indispensable for understanding the underlying causes of the differences in spectroscopic and photometric properties and thus for improving the current methods for deriving precise extragalactic distances.

In this work, we pursue the recent effort to do statistical analyses of large samples to better understand SN~II diversity. For this purpose, we use photometric and spectroscopic observations of 55 local SNe~II obtained by the UC Berkeley SN group. During the past two decades and under the Lick Observatory Supernova Search \citep[LOSS]{filippenko01,leaman11}, the UC Berkeley SN group has been one of the most active groups in SN discoveries ($\sim40$\% of nearby SNe during the years 1998--2008; a smaller fraction thereafter, with the advent of wide-angle SN surveys). Their efforts have permitted the building of large datasets of any type of SN and led to the publications of a wide range of studies: SNe~Ia \citep{ganeshalingam10,silverman12,silverman12b,silverman12c,silverman12d,zheng17}, SN rates \citep{li2011,shivvers17,graur17a,graur17b}, stripped-envelope supernovae \citep{matheson01,shivvers19}, SNe~IIn \citep{bilinski15}, and SN~II \citep{poznanski09,poznanski10,faran14a,faran14b,silverman17}. However, even if a few individual objects have been published recently in the literature, such as SN~2009kr \citep{eliasrosa10}, SN~2010id \citep{galyam10}, and SN~2013ej \citep{dhungana16}, not all of the SN~II photometric and spectroscopic data gathered by the UC Berkeley SN group since \citet{faran14a,faran14b} have been published.

Therefore, the purpose of this paper is to report SN~II photometric and spectroscopic data obtained by our group since 2005 and the last SN~II Berkeley data sample release by \citet{faran14a,faran14b}. All the dataset will be immediately available to the community and the reader can find information on each SN in Appendix \ref{AppendixA}, Table \ref{SN_sample}. Note that this paper is part of a more extensive data release, including stripped-envelope SNe (Zheng et al., in prep.) and SNe~Ia (Stahl et al., in prep.). 

This paper is divided as follows. Section \ref{txt:data_sample} describes optical observations and data-reduction procedures; Section \ref{txt:Results} presents an analysis of the photometric and spectroscopic properties of our sample, including light curves, colours, absolute magnitudes, velocities, and time evolution of spectral lines. Finally, Section \ref{txt:conclusions} contains a summary and the conclusions.

\section{Data Sample}\label{txt:data_sample}

The Berkeley SN~II sample consists of 55 objects observed between 2006 and 2018 using the Lick Observatory (Mt. Hamilton, CA) facilities. Among these transients, 30 were discovered by LOSS \citep{filippenko01}. For almost all the SNe, spectra were obtained using the 3~m Shane Lick telescope and the Keck-I/Keck-II 10~m telescopes in Hawaii (see Section \ref{txt:data_spec}). However, for 9 SNe~II \footnote{SN~2007il, SN~2009ao, SN~2012ec, SN~2013bu, SN~2013ft, SN~2014dq, SN~2016X, SN~2016cyx, and SN~2017jbj} our group did not obtain any spectra, and therefore we complete our spectroscopic sample with spectra available in the literature. 

\begin{figure}
\includegraphics[width=1.0\columnwidth]{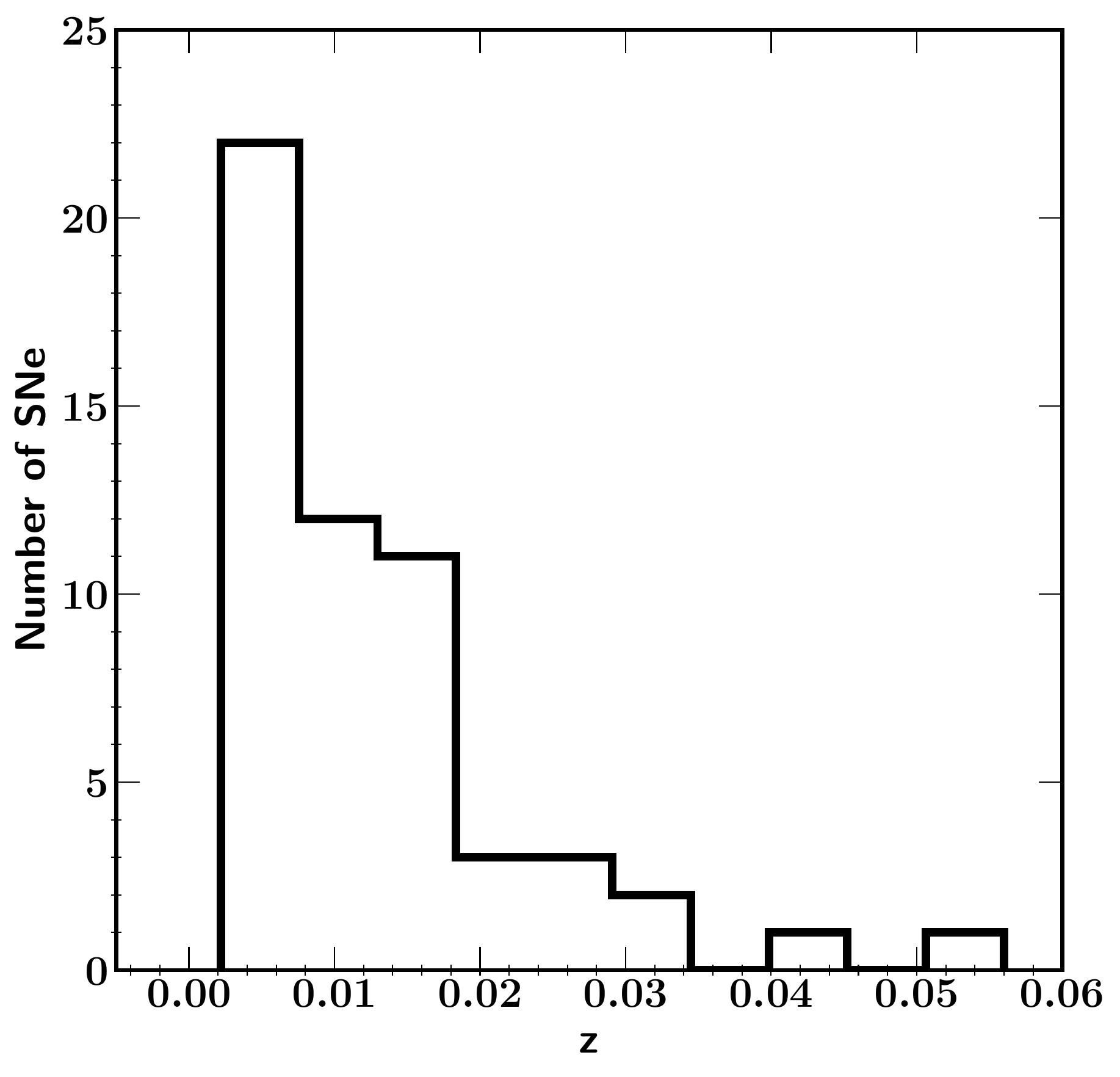}
\caption{In this figure shows the redshift distribution of the 55 SNe~II. The average value of the
distribution is 0.0125 with a standard deviation of 0.0103. 29 SNe~II have $z>0.01$.}
\label{fig:z_distribution}
\end{figure}

The heliocentric redshifts were obtained from the host-galaxy recession velocities published in the NASA/IPAC extragalactic Database (NED\footnote{\url{http://ned.ipac.caltech.edu/}}) when available, otherwise from the SN spectra. The redshift distribution of the Berkeley SN~II sample is presented in Figure \ref{fig:z_distribution}; the redshift ranges from 0.0022 (SN~2013ej) to 0.0559 (SN~2015O) with an average value of 0.0125 and a standard dispersion of 0.0103. Note that 29 SNe~II are located in the Hubble flow ($z>0.01$). In Appendix \ref{AppendixA}, Table \ref{SN_sample}, the reader can find information on each SN: its host galaxy, dust extinction from the Milky Way (MW), recession velocity, distance modulus, explosion epoch, last nondetection and detection epochs, number of photometric points and number of spectra. 

For each SN, to determine the explosion date, the same methodology used by \citet{anderson14a} or \citet{galbany16a} was applied. When nondetections are available, the explosion date is taken as the intermediate epoch between the last nondetection and the first detection, and its uncertainty corresponds to half of this duration. When nondetections are not available, the explosion date is obtained using SNID \citep{blondin07} by matching SN~II spectral templates with well-constrained explosion epochs. The explosion date is then taken as the average epoch of the best fits and its uncertainty as the standard deviation \citep{gutierrez17b}. Note that three SNe~II (SN~2016bkv, SN~2018aoq, SN~2018bek) observed by our group will be published in more detailed studies (Van Dyk et al., in prep.; Lymin et al., in prep.). 

\subsection{Photometry}\label{txt:data_phot}

SN images were obtained using the 0.76~m Katzman Automatic Imaging Telescope (KAIT) and the 1~m Nickel telescope, both at Lick Observatory (average seeing $\lesssim2$\arcsec). The majority of our images ($\sim 65$\%) were taken with the completely robotic KAIT telescope and an exposure of 60~s in $BVRI$, while the exposure times for the Nickel images average 600~s and 300~s for $B$ and $VRI$, respectively. For more information concerning the transmission curves and the colour terms for the KAIT and Nickel telescopes, the readers are referred to Stahl et al. 2019 (in prep.). 

The photometric reductions are fully described by Stahl et al. 2019 (in prep.); here we only briefly summarise the procedure. Using our automated image-reduction pipeline (\citealt{ganeshalingam10} and Stahl et al. 2019 (in prep.)), we applied to all of the images bias removal, flat-field corrections, and astrometric solution. A majority of SNe ($\sim 60$\%) were relatively close from their host galaxy, and therefore, to remove the host-galaxy luminosity, galaxy subtraction were required. Subtraction templates were obtained on a dark night using the Nickel telescope and after the SN had faded beyond detection (generally at least 1~yr  after the discovery).

Finally, using DAOPHOT \citep{stetson87} from the IDL Astronomy User's Library, point-spread-function (PSF) photometry was performed to to measure the SN flux relative to local standard stars in the same field. Instrumental magnitudes were calibrated using two or more standard stars (depending on the field) from the Pan-STARRS1 Surveys \citep[PS1]{chambers16,schlafly12}. ‎$grizy$ PS1 magnitudes were transformed into the Landolt standard system \citep{landolt92} using the transformations given by \citet{tonry12}. Finally, the transformation between the standard Landolt system into instrumental magnitudes was achieved using the following equations:

\begin{subequations}
\begin{align}
b &= B + C_{B}(B-V) + {\rm constant},\\
v &= V + C_{V}(B-V) + {\rm constant},\\
r &= R + C_{R}(V-R) + {\rm constant},~{\rm and}\\
i &= I + C_{I}(V-I) + {\rm constant},
\end{align}
\end{subequations}
\noindent
where lower-case and upper-case bandpass letters are (respectively) the instrumental magnitudes and the Landolt magnitudes. The coefficient $C_{i}$ ($i=B,V,R,I$) represent the average colour terms published by \citet{ganeshalingam10} and Stahl et al. 2019 (in prep.). Note that there are no atmospheric effects or zero points, as they are absorbed into the constant. Finally, it is worth noting that the SN~II photometry is released in the natural system of the KAIT and Nickel telescopes (transmission curves are available in Stahl et al. 2019 in prep.).

\subsection{Spectroscopy}\label{txt:data_spec}

Optical spectra were obtained using the Kast double spectrograph \citep{miller93} on the 3~m Shane telescope at Lick Observatory (155/213 spectra), the Low Resolution Imaging Spectrometer (LRIS; \citealt{oke95}) mounted on the Keck-I 10~m telescope (16/213 spectra) located on Maunakea (Hawaii), and the DEep Imaging Multi-Object Spectrograph (DEIMOS \citealt{faber03}) mounted on Keck-II 10~m telescope also located on Maunakea (7/213 spectra). To minimise light losses due to atmospheric dispersion, all of the specta were obtained at (or near) the parallactic angle \citep{filippenko82}. 

To reduce our spectroscopic data, we use two fully automated public pipelines: TheKastShiv\footnote{\url{https://github.com/ishivvers/TheKastShiv}} for Kast spectra and LPIPE\footnote{\url{http://www.astro.caltech.edu/~dperley/programs/lpipe.html}} to reduce LRIS longslit spectrum \citep{perley19}. Briefly, these two pipelines follow standard spectroscopic reduction techniques. First, the spectra were debiased, flat-field, and cleaned of cosmic rays. Then, one dimensional spectra were extracted and calibrated using lamps. Finally, spectrophotometric standard stars observed on the same night are used to calibrate the flux and removed atmospheric absorption lines.


To complete the spectral analysis, 35 spectra from the literature were added to our sample. Of these 35 spectra, four (of SN~2013bu, SN~2015X, SN~2017jbj, and SN~2016cyx) were unpublished but publicly available and were downloaded from the WISeREP database\footnote{\url{https://wiserep.weizmann.ac.il/}}; the others were obtained from the WISeREP database or from electronic links in the published manuscripts. These 35 spectra were obtained with the 2.5~m Ir\'en\'ee du Pont telescope using the WFCCD and the Boller and Chivens spectrographs and the 6.5~m Magellan Clay and Baade telescopes with LDSS-2 and LDSS-3 at Las Campanas Observatory (SN~2007il, SN~2009ao; \citealt{gutierrez17b}), the Australian National University 2.3~m telescope with the Wide-Field Spectrograph (SN~2012ec, SN~2014dq; \citealt{childress16}), the 1.82~m telescope at Cima Ekar with the AFOSC spectrograph (SN~2013bu, SN~2015X, SN~2017jbj), the 2.56~m Nordic Optical Telescope with the ALFOSC instrument (SN~2013ft; \citealt{khazov16}), and the Fred Lawrence Whipple Observatory 1.5~m telescope with the FAST spectrograph (SN~2016cyx). The majority of our spectra cover a wavelength range of 3600--10,000~\AA\ with a resolution of $\sim 10$~\AA.

\begin{figure}
\includegraphics[width=1.0\columnwidth]{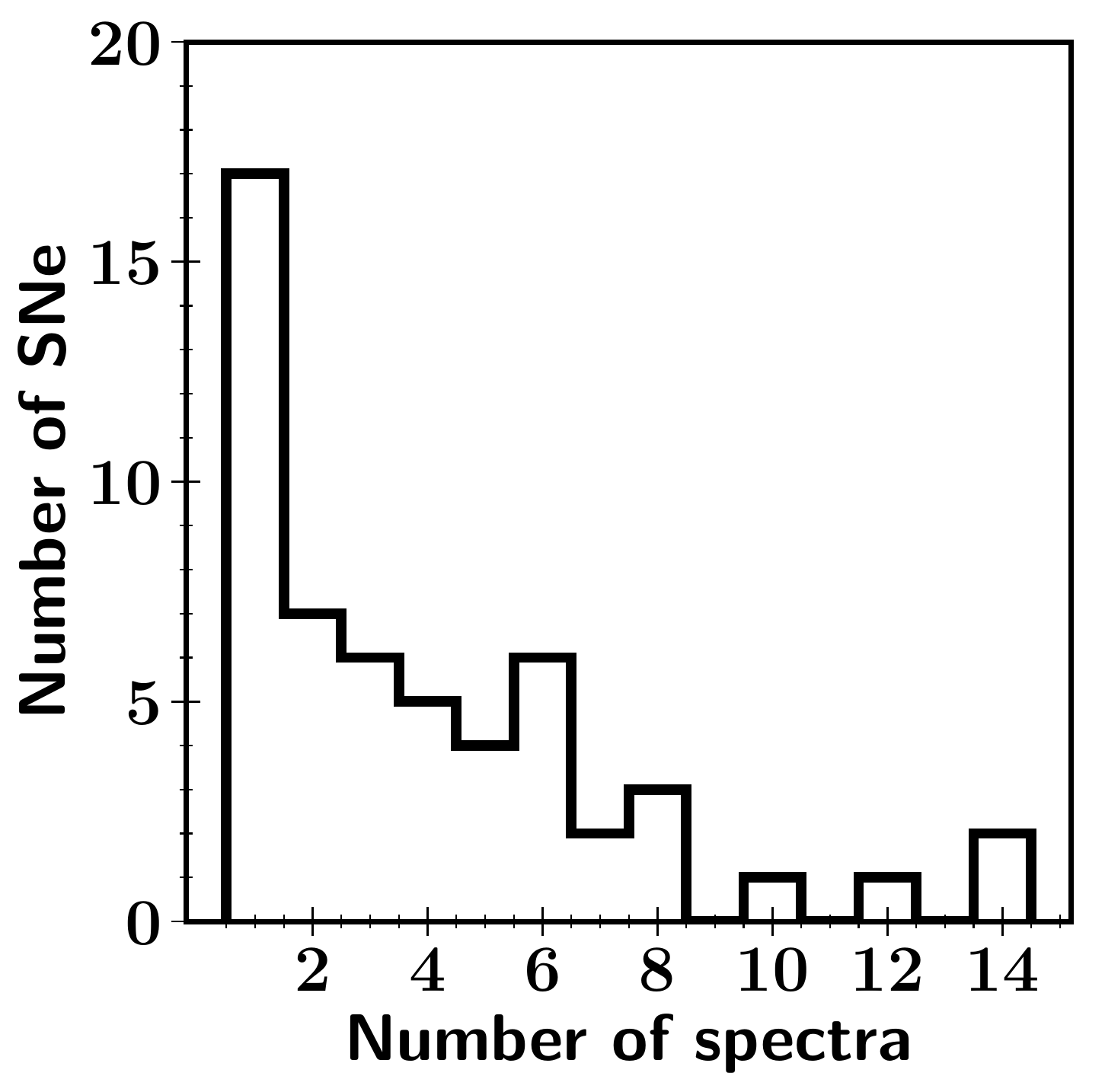}
\caption{Black histogram represents the number of spectra per SN. In total, our sample is composed of 213 spectra and 55 SNe~II. }
\label{fig:spec_SN_distribution}
\end{figure}

The distribution of the number of spectra per object shown in Figure \ref{fig:spec_SN_distribution} peaks at one spectrum per SN and has a median value of three spectra. 17 SNe only have one spectrum, one SN has no spectrum (SN~2014cn), and $\sim 70$\% of the SNe in our sample have at least two spectra. SN~2013ab and SN~2015V are the SNe with the most spectra (14), and 15 SNe have more than five spectra each.

\section{Results}\label{txt:Results}

In this Section, we present the photometric properties (light and colour curves, absolute peak magnitudes, and slopes) and spectroscopic properties (velocities, pseudoequivalent widths) of our sample. All of these characteristics are also compared to the low-$z$ SN~II samples published in the literature \citep{anderson14a,galbany16a,gutierrez17b}.

\subsection{Photometric analysis}

\subsubsection{Light curves}
In Figure \ref{fig:LC_KAIT}, we present 55 $BVRI$ and Clear (i.e., unfiltered) light curves in the natural KAIT/Nickel photometric system. All magnitudes have been corrected for MW extinction using the dust maps of \citealt{schlafly11} assuming $R_{V}=3.1$ and the \citet{car89} extinction law. We decided to not correct for host-galaxy extinction as to date no accurate methods exist to estimate it \citep{poznanski11,phillips13,faran14a,galbany17,dejaeger18a}. Neither $K$-corrections \citep{oke68,hamuy93,kim96} nor $S$-corrections \citep{stritzinger02} have been applied owing to the low redshifts of our objects (see Figure \ref{fig:z_distribution}) and the similarity between the different filters of the KAIT/Nickel system.

\begin{figure}
\includegraphics[width=1.0\columnwidth]{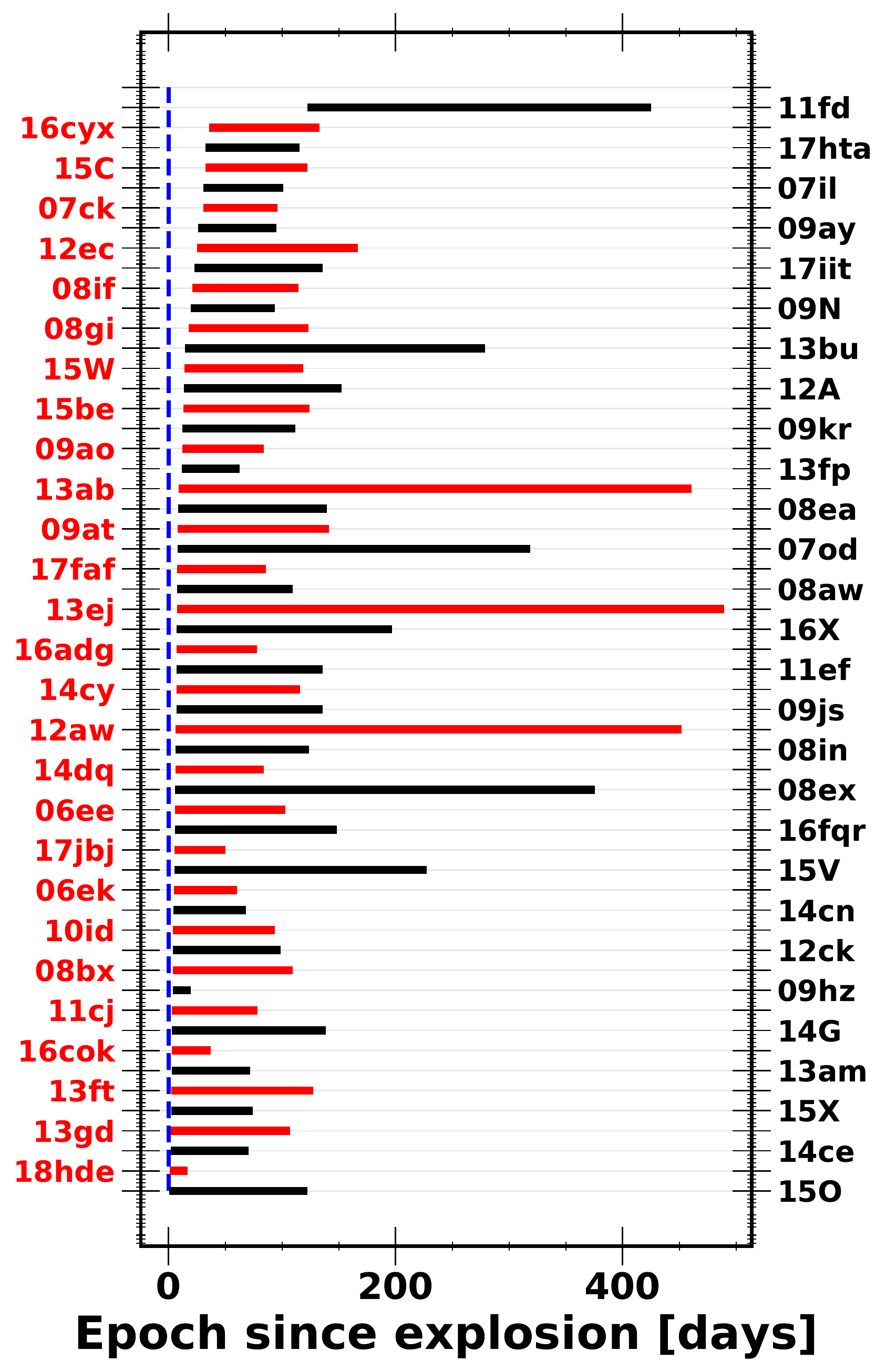}
\caption{$BVRI$ light-curve coverage for each SN sorted by increasing first photometric observation. The vertical blue dashed line represents the explosion date.}
\label{fig:coverage_LC}
\end{figure}

In Figure \ref{fig:coverage_LC}, the light-curve coverage for each SN is shown. The photometric observations start on average 12 days after the explosion with a standard deviation of 17 days. 16 SNe have their first photometric point before five days since the explosion and the vast majority of the objects before 10 days (66\%). On average, the last optical images were obtained 144 days after the explosion with a standard deviation of 110 days. Two thirds of the SNe have photometric data $>100$ days after the explosion. Each SN has an average of 93 photometric points with a standard deviation of 77 points, and almost half of the SNe have at least 80 points. With 449 photometric points, SN~2013ej is the object with the best photometric coverage, followed by SN~2015V with 287 points. For SN~2006ek, we obtained only 12 optical images, our poorest sampling. The total photometric points published in this work is 5093.

\subsubsection{Colour curves}\label{txt:colour-curve}
\begin{figure*}
\includegraphics[width=0.8\textwidth]{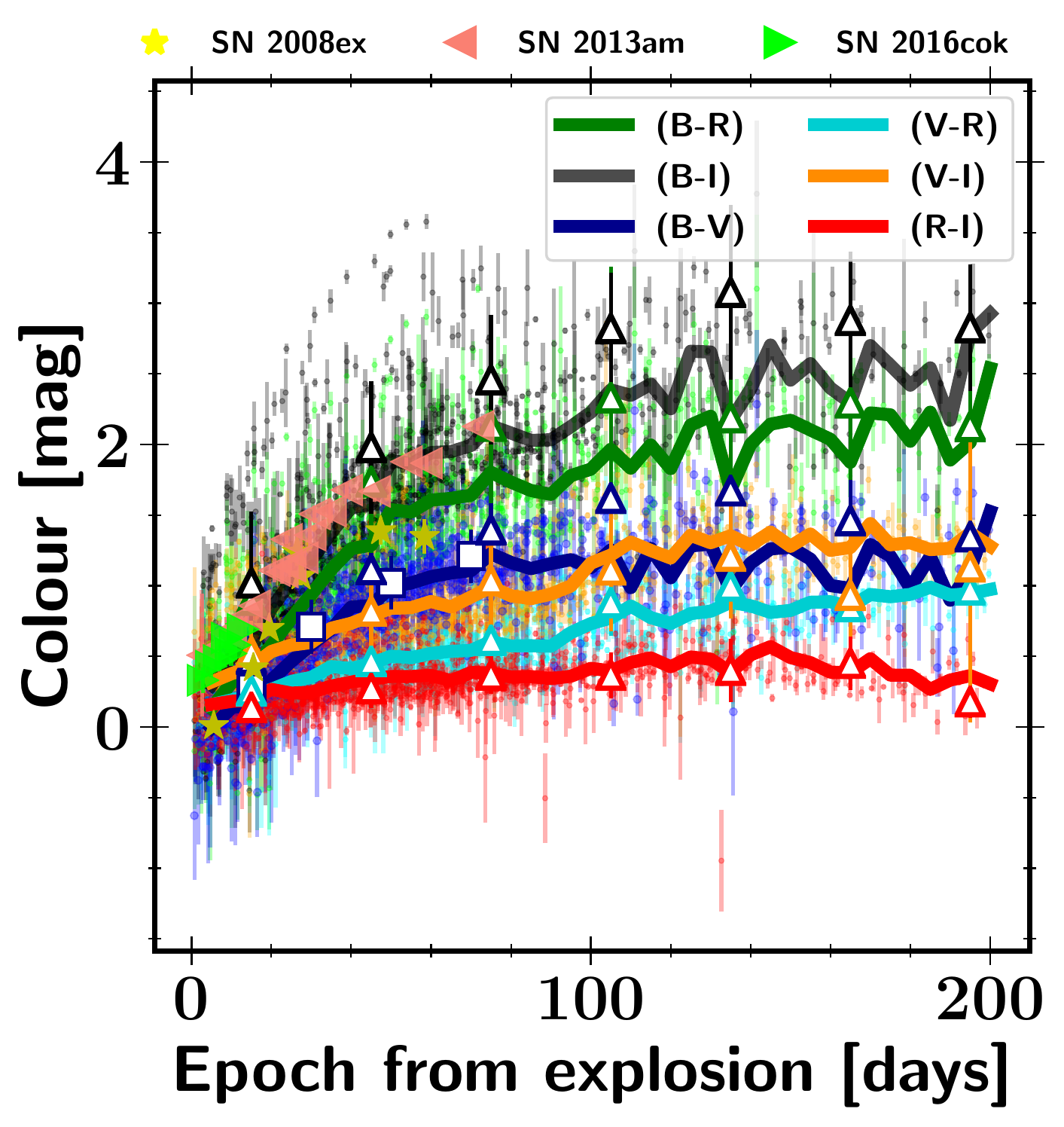}
\caption{55 $(B-R)$, $(B-I)$, $(B-V)$, $(V-I)$, $(V-R)$, and $(R-I)$ colour evolution corrected for MW extinction are respectively represented in green, black, blue, orange, cyan, and red. Individual measurements are shown with dots while solid lines indicate average colours in a bin size of five days. Blue squares represent the $(B-V)$ colour at 15, 30, 50, and 70 days after the explosion from the Carnegie Supernova Project-I \citep{dejaeger18a}. Empty triangles represent the average colour in a bin size of 30 days published by \citet{galbany16a}. The $(B-V)$ colours of SN~2008ex, SN~2013am, and SN~2016cok are respectively highlighted using yellow stars, salmon left-pointed triangles, and lime right-pointed triangles.}
\label{fig:colour_evolution}
\end{figure*}

In Figure \ref{fig:colour_evolution}, six different colour curves of the 55 SNe~II are represented ($(B-V)$, $(B-R)$, $(B-I)$, $(V-R)$, $(V-I)$, and $(R-I)$). As expected, all of the colours follow the general SN~II colour behaviour: at early times (30--40 days) a rapid increase is seen while at later the increase is much slower. Finally, at late epochs ($>80$--100 days) the colour curves are flatter, as they all depend on the $^{56}$Co decay \citep{galbany16a}. Differences in colour evolution between the different colours are also seen. The redder colours increase more slowly than the bluer colours because the red part of the spectrum is less sensitive to temperature changes than the blue part. With our LOSS sample, we do not see two distinct patterns of colour evolution in any of the colour curves, and therefore, confirming that SNe~II form an unique class \citep{anderson14a,sanders15,valenti16,galbany16a,dejaeger18a}.

As described by \citet{galbany16a}, all of the colours do not show the same dispersion, with $(B-I)$ showing the largest scatter while $(R-I)$ the smallest. \citet{galbany16a} attributed this scatter to host-galaxy dust as reddening is strongest at bluer bands. However, \citet{dejaeger18a} have shown that host-galaxy extinction does not seem to be the principal parameter to explain the dispersion in observed colours. They suggested that the main parameter affecting the observed colour diversity is intrinsic, depending on differences in progenitor radius and/or circumstellar material around the progenitor stars.

Even if the majority of the SN~II colours diversity is intrinsic, the reddest SNe~II should be affected by host-galaxy extinction. In our sample, we identify two objects (SN~2013am and SN~2016cok) whose $(B-V)$ colour differs by $>2\sigma$ from the average colour. These objects are highly extinguished: $\sim2$ mag for SN~2013am \citep{tomasella18} and $\sim1.5$ mag for SN~2016cok \citep{kochanek17}. Note also that SN~2008ex shows red colours that can be explained by an unusual SN~II optical light curve (after cooling the brightness increases).

Finally, following \citet{dejaeger18a}, as $(B-V)$ colour curves can be described with one or two slopes, we perform for each SN a weighted least-squares fit of the $(B-V)$ colour curves. From our sample, the first colour slope, the second slope, and the epoch of transition have average values ($N=16$) of 2.52 $\pm$ 0.50, 0.48 $\pm$ 0.19, and 38.2 $\pm$ 5.62, respectively. These values are consistent with those published by \citet{dejaeger18a}: 2.63 $\pm$ 0.62, 0.77 $\pm$ 0.25, and 37.7 $\pm$ 4.31. Similarly, the $(B-V)$ values at 15, 30, 50, and 70 days after the explosion are also consistent with those derived by \citet{dejaeger18a}: 0.30 $\pm$ 0.21, 0.69 $\pm$ 0.24, 0.96 $\pm$ 0.25, and 1.06 $\pm$ 0.29, respectively. However, with this sample, we do not recover the correlation found by \citet{dejaeger18a} between the first and second slope. Absence of a statistically significant trend is explained by the small number of objects with good temporal coverage to see two slopes ($N=16$). To address the small number statistics issue, we add to our sample 28 SNe~II from the previous LOSS SN II data release \citep{faran14a,faran14b}. All these SNe~II have well defined explosion dates and were observed under the same conditions (same telescopes, same pipeline). From this new sample, 14 SNe~II have enough temporal coverage to see two slopes and therefore, the total SN II number increases to 30 SNe~II ($N=16 +14$). With this new sample, the correlation found by \citet{dejaeger18a} between the first and second colour slopes is confirmed with a Pearson factor of 0.54 $\pm$ 0.14 ($p \leq 2.0 \times 10^{-2}$).

\subsubsection{Absolute magnitudes}\label{mag_max}

For each SN, absolute magnitudes are calculated using the distance modulus and the SN apparent magnitudes corrected only for MW extinction. The distance modulus is obtained using the cosmic microwave background corrected recession velocities if the value is higher than 3000 km s$^{-1}$ and assuming a $\Lambda$CDM model ($\Omega_{m}=0.3$, $\Omega_{\Lambda} = 0.7$) with a Hubble constant of 70 km s$^{-1}$ Mpc$^{-1}$. An uncertainty of 300 km s$^{-1}$ is added to take into account the galaxy peculiar velocities. For recession velocities smaller than 3000 km s$^{-1}$, peculiar-motion errors are too large making the distance measurement unreliable. For these cases, following \citet{anderson14a}, the distance moduli are collected from NED (see Table \ref{SN_sample}) and based Cepheids, Tully-Fisher relation, or SN~II methods.

\begin{figure}
\includegraphics[width=1.0\columnwidth]{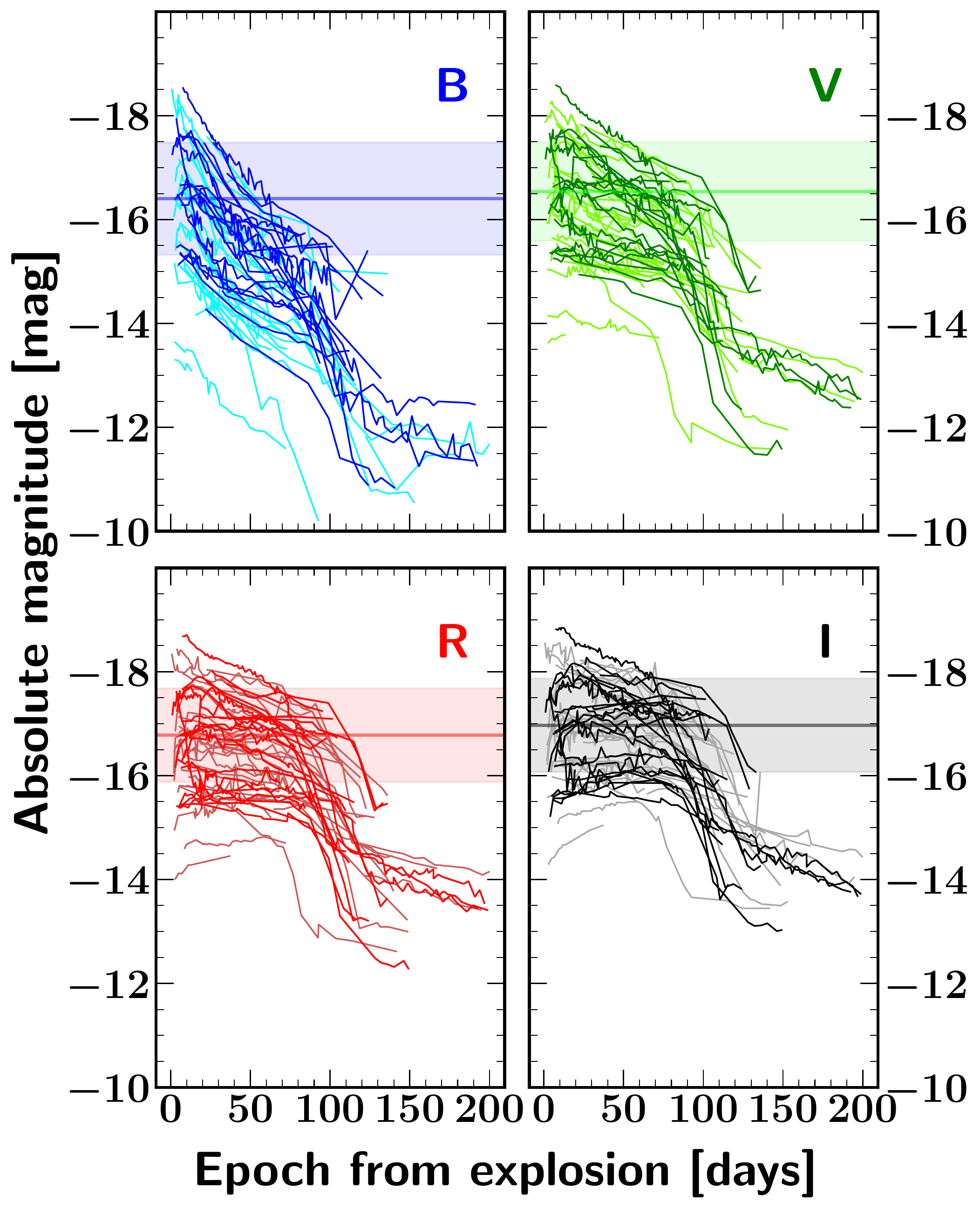}
\caption{55 $B$, $V$, $R$, and $I$ absolute magnitude light curves are displayed in four different panels. All the light curves have been interpolated using zero-order spline polynomials. In each panel, the darkest colours represent the SNe with the bluest $(B-V)$ colour. The horizontal line and the filled region represent the average peak magnitudes and their 1$\sigma$ uncertainties, respectively.}
\label{fig:mag_abs}
\end{figure}

In Figure \ref{fig:mag_abs}, for each SN, the $BVRI$ absolute magnitude light curves are displayed in separate panels. We see that the absolute peak magnitudes spread over a wide range of $-14$ to $-18.5$ mag. The two brightest objects are SN~2017faf and SN~2012ck ($M_V \approx -18.5$ mag and $M_V \approx -18.2$ mag), while SN~2013am and SN~2016cok are the faintest objects in our sample ($\sim -14$ mag). However, as mentioned in Section \ref{txt:colour-curve}, these two objects are highly extinguished \citep{tomasella18,kochanek17}. If we restrict our sample to the bluest objects by selecting only the SNe having an average $(B-V)$ colour less than the average $(B-V)$ colour of the whole sample, the range of absolute peak magnitude is still large ($-15$ to $-18.5$ mag). This suggests that the absolute magnitude range has an intrinsic origin in lieu of host-galaxy extinction. Note also that as demonstrated by \citet{anderson14a} for the $V$ band and confirmed later by \citet{sanders15,galbany16a} for all the bands, even if our sample shows a wide range in both absolute magnitudes and light-curve morphologies, there is no evidence of two separate subpopulations, confirming that the SN~II progenitor originates from a single stellar population.

Finally, for each SN and each filter, we derive the absolute magnitude at peak brightness using a low-order polynomial fit to the photometry close to the maximum photometric point. Otherwise, for the majority of the cases, due to the absence of peak (lack of early data), the maximum brightness is taken as the first photometric point if the epoch is less than 20 days post-explosion. Our average absolute peak magnitudes excluding SN~2013am and SN~2016cok (two highly extinguished SNe) are $<B_{\rm max}>=-16.39$ mag ($\sigma=1.08$, $N=42$), $<V_{\rm max}>=-16.53$ mag ($\sigma=0.94$, $N=42$), $<R_{\rm max}>=-16.74$ mag ($\sigma=0.94$, $N=41$), and $<I_{\rm max}>=-16.95$ mag ($\sigma=0.89$, $N=42$). These values are slightly lower ($\sim 0.2$--0.3 mag) than those published by \citet{galbany16a} but are still consistent within the uncertainties ($<B_{\rm max}>=-16.43$ mag, $<V_{\rm max}>=-16$.89 mag, $<R_{\rm max}>=-16.96$ mag, and $<I_{\rm max}>=-17.27$ mag). 

The small differences above can be explained mostly by (1) the uncertainties in the SN distances (almost 20 SNe~II have distances from the Tully-Fisher relation), (2) the fact that no clear maximum is seen, and therefore the first photometric point is only an approximation of the maximum, (3) observational selection effects (KAIT targets bright galaxies; \citealt{leaman11}), and (4) the uncertainties added by the host-galaxy extinction into the absolute peak magnitude values. For example, for the $V$ band, only 10 SNe~II have their maximum derived from a polynomial fit. If we select only those SNe, the average absolute peak magnitudes is brighter ($-17.17$ mag). Now, if we derive the average peak magnitude only for the SNe having a recession velocity higher than 3000 km s$^{-1}$, the new values obtained are more consistent with those of \citet{galbany16a}: $<B_{\rm max}>=-16.57$ mag ($\sigma=1.14$, $N=23$), $<V_{\rm max}>=-16.74$ mag ($\sigma=0.92$, $N=23$), $<R_{\rm max}>=-16.96$ mag ($\sigma=0.85$, $N=23$), and $<I_{\rm max}>=-17.20$ mag ($\sigma=0.82$, $N=23$). Note that among the SNe having the lowest absolute peak magnitudes ($V$ band), two objects are highly extinguished (SN~2013am and SN~2016cok), while others have already been discussed in the literature such as SN~2008in \citep{roy11}, SN~2009N \citep{takats14}, and SN~2010id \citep{galyam10}.

\subsubsection{Light-curve properties}

In this section, following \citet{anderson14a} and \citet{galbany16a}, we investigate the SN~II light-curve properties by measuring two different parameters: (1) the decline rate in magnitudes per 100 days between the peak brightness and the end of the plateau and (2) the optically thick phase duration (OPTd) which is equivalent to the epoch of the end of the plateau phase. 
 
For our sample, the average plateau length in the $V$ band is 86 $\pm$ 11 days, similar to those published by \citet{anderson14a} and \citet{galbany16a}: 83.7 $\pm$ 16.7 days and 77.5 $\pm$ 26.3 days, respectively. With a duration of 60 days, SN~2017faf has the shortest plateau, while SN~2014cy with a duration of 104 days has the largest OPTd. For $s_{1}$, $s_{2}$, $s_{3}$, and $s$ we derive respective average values of 2.60 mag (100 days)$^{-1}$ ($\sigma=1.10$; $N=10$), 1.29 mag (100 days)$^{-1}$ ($\sigma=0.90$; $N=45$), 1.15 mag (100 days)$^{-1}$ ($\sigma=0.35$; $N=11$), and 1.38 mag (100 days)$^{-1}$ ($\sigma=0.91$; $N=45$). These values are also consistent with those published by \citet{anderson14a} [2.65 mag (100 days)$^{-1}$ ($\sigma=1.50$; $N=28$), 1.27 mag (100 days)$^{-1}$ ($\sigma=0.93$; $N=113$), and 1.47 mag (100 days)$^{-1}$ ($\sigma=0.82$; $N=30$)] and by \citet{galbany16a} [1.53 mag (100 days)$^{-1}$ ($\sigma=0.91$; $N=45$)], respectively. 

\begin{figure}
\includegraphics[width=1.0\columnwidth]{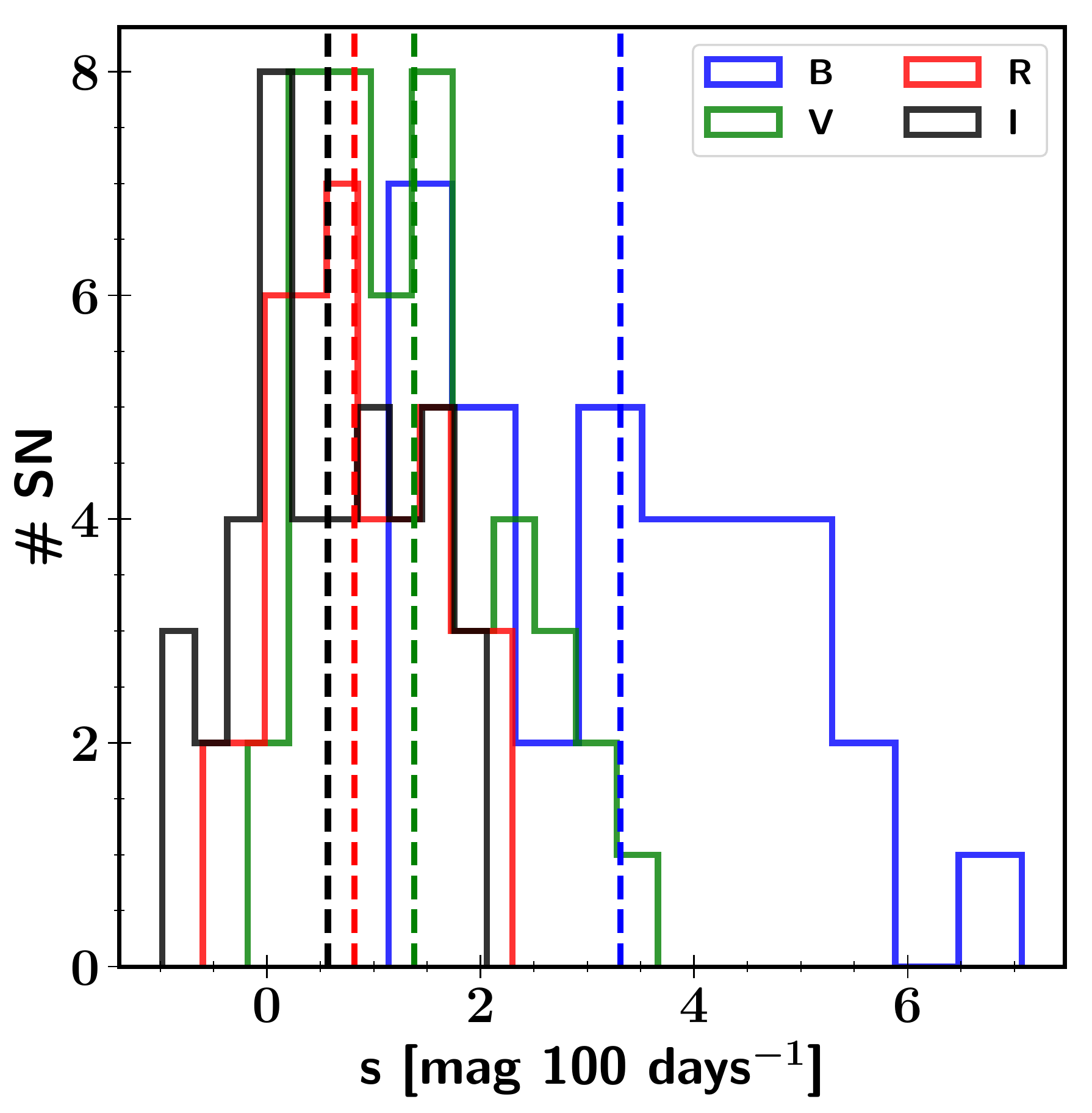}
\caption{Histograms of the $s$-parameter distributions in each band for the Berkeley SN~II sample. The vertical dashed line represents the average value. $B$, $V$, $R$, and $I$ histograms are displayed in blue, green, red, and black, respectively.}
\label{fig:s_distri}
\end{figure}

In Figure \ref{fig:s_distri}, histograms of the $s$-parameter distributions in each band are displayed. As expected, a trend is seen between the filter and the decline rate, in the sense of SNe~II declining more steeply in bluer bands than in redder bands. For each band we derive an average decline rate of 3.31 mag (100 days)$^{-1}$ ($\sigma=1.49$; $N=34$), 1.38 mag (100 days)$^{-1}$ ($\sigma=0.91$; $N=45$), 0.82 mag (100 days)$^{-1}$ ($\sigma=0.73$; $N=42$), 0.57 mag (100 days)$^{-1}$ ($\sigma=0.81$; $N=45$) for $B$, $V$, $R$, and $I$, respectively. These values are similar with those published by \citet{galbany16a}. In contrast to the $s$ parameter, the OPTd values are similar for all bands, with only a slight (but not significant) increase for redder bands (85 $\pm$ 14, 86 $\pm$ 11, 87 $\pm$ 10, and 87 $\pm$ 11 days for $B$, $V$, $R$, and $I$, respectively). Note also, if we add the previous Berkeley SN~II data release \citep{faran14a,faran14b} to our sample, the distributions and the average values are almost identical.

\subsubsection{Brightness and decline-rate correlations}\label{txt:s_corr}

In this section, we investigate the correlation derived by various authors \citep{anderson14a,sanders15,Pejcha15b,valenti16,galbany16a} between the absolute peak magnitude and the plateau slope --- that is, rapidly declining SNe~II are generally more luminous than slowly declining SNe~II. This relation has also been used to standardise SNe~II and to derive extragalactic distances with a precision of $
\sim 18$\% \citep{dejaeger15b,dejaeger17a}.

\begin{figure}
\includegraphics[width=1.0\columnwidth]{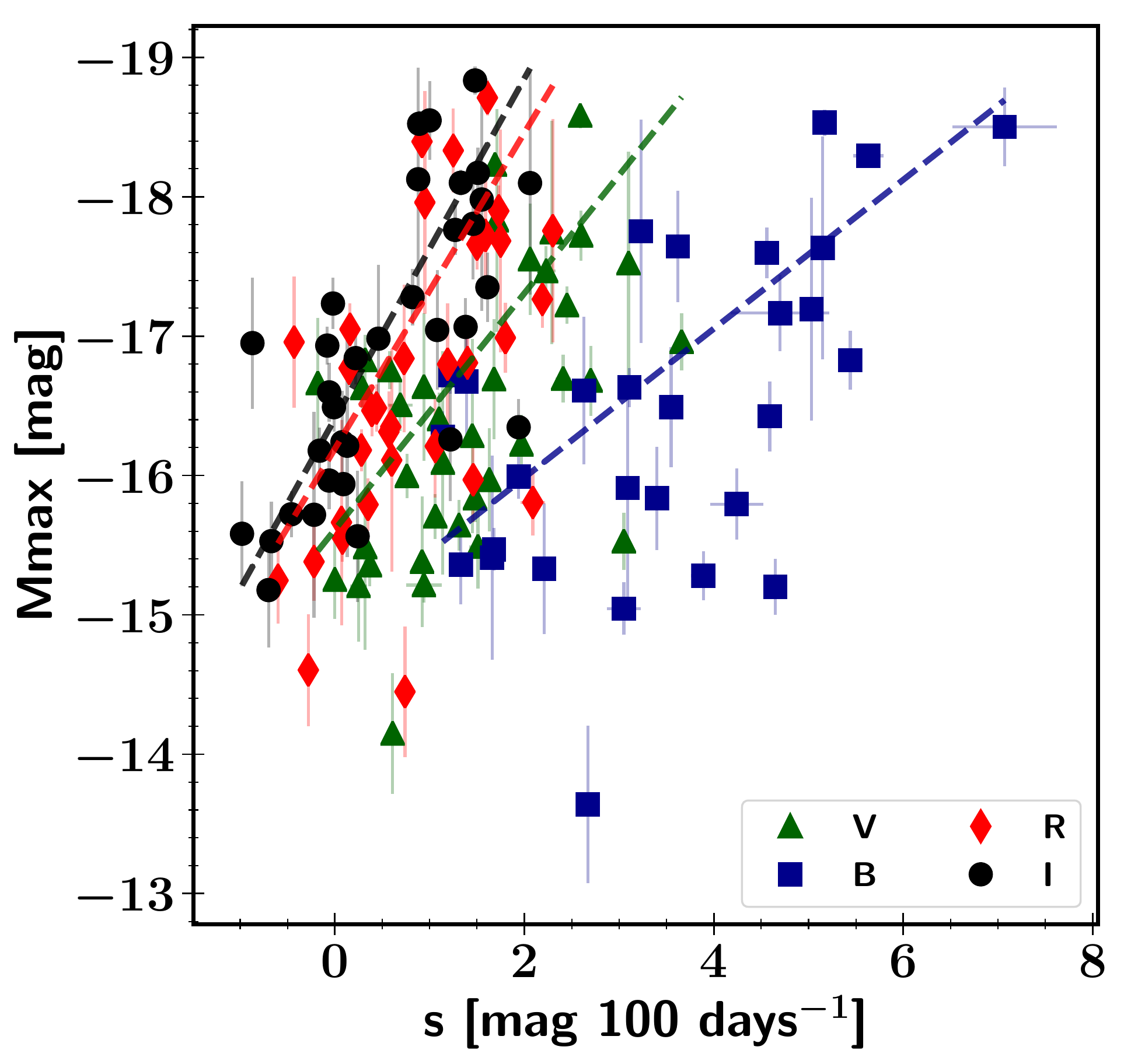}
\caption{Absolute peak magnitudes versus the slope in mag per 100 days between the maximum brightness and the end of the plateau ($s$ parameter). Blue squares, green triangles, red diamonds, and black circles represent $B$, $V$, $R$, and $I$, respectively. In all the bands, more-luminous SNe~II have steeper decline rates.}
\label{fig:Mmax_s}
\end{figure}

In Figure \ref{fig:Mmax_s}, the absolute peak magnitude ($M$max) versus the decline rate between the maximum brightness and the end of the plateau ($s$ parameter) is plotted. In all the bands, a statistically significant correlation is seen between those two quantities, i.e., brighter SNe~II decline faster. The average Pearson factors are $r_{B} = -0.56 \pm 0.12$ ($N=29$, $p \leq 4.0 \times 10^{-2}$), $r_{V} = -0.53 \pm 0.11$ ($N=37$, $p \leq 9.0 \times 10^{-3}$), $r_{R} = -0.61 \pm 0.11$ ($N=33$, $p \leq 3.5 \times 10^{-3}$), and $r_{I} = -0.72 \pm 0.09$ ($N=34$, $p \leq 1.0 \times 10^{-4}$). Our results also support the existence of the correlation between $s$ and $M$max for the $B$ band \citep{galbany16a}, contrary to \citet{Pejcha15b} who do not find a correlation for bands with $\lambda <0.5\,\mu$m.

\begin{figure}
\includegraphics[width=1.0\columnwidth]{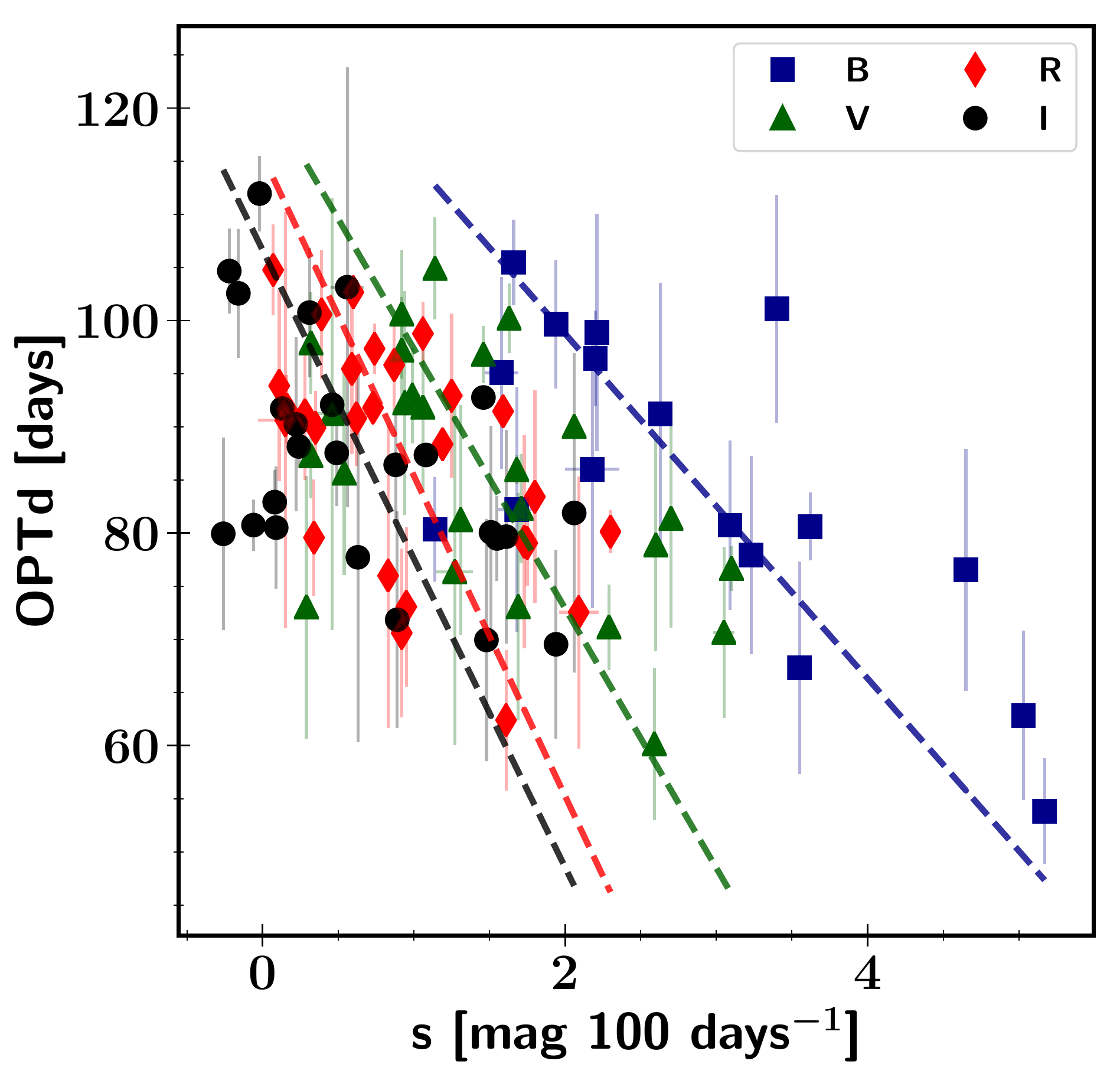}
\caption{Optically thick duration (OPTd; plateau phase) versus decline rate in mag per 100 days between the maximum brightness and the end of the plateau ($s$ parameter). Blue squares, green triangles, red diamonds, and black circles represent $B$, $V$, $R$, and $I$, respectively. In all the bands, faster-declining SNe~II have shorter OPTd.}
\label{fig:OPTd_s}
\end{figure}

Figure \ref{fig:OPTd_s} shows the correlation between the OPTd and the decline rate. In all the bands, the OPTd distribution ranges from $\sim 60$ to $\sim 110$ days with an average value of 86 days. From this figure, we see that SNe~II with steeper decline generally have shorter plateau duration. The average slope of this correlation is $-24.964 \pm 5.481$, once again consistent with the value derived by \citet{galbany16a}. However, contrary to their work, the strength of the correlation does not increase from bluer to redder bands and remains mostly similar in each band ($Pearson factor \pm 0.5$--0.6).

Finally, these two correlations between the decline rate and the absolute magnitude or the OPTd agree with previous observational and theoretical work \citep{blinnikov93,popov93,anderson14a,sanders15,Pejcha15b,valenti16,galbany16a}. SN progenitors with smaller hydrogen envelopes have shorter OPTd because the radiation is trapped for a shorter time and are brighter as a larger fraction of radiation can escape. It is worth noting that the narrow plateau duration distribution ($\sim 15$\%) could be explained by the idea that if the energy varies directly as the cube of the mass, the plateau duration only depends slightly on the radius \citep{poznanski13}.

\subsection{Spectroscopic analysis}

At early phases, a SN~II spectrum exhibits a blue continuum (10,000 K) with strong P-Cygni profiles of Balmer lines (H$\alpha$ $\lambda 6563$, H$\beta$ $\lambda 4861$, H$\gamma$ $\lambda 4341$) and the He~I $\lambda 5876$ line. With time, the SN ejecta will expand and the inner products will start to appear as (for example) \ion{Fe}{II} $\lambda\lambda 4924$, 5018, 5169, \ion{Na}{I}~D $\lambda\lambda 5890$, 5896, \ion{O}{I} $\lambda 7774$, or \ion{Ca}{II} $\lambda\lambda 8498$, 8542, 8662 (also \ion{Sc}{II}, \ion{Ba}{II}, \ion{Ti}{II}; see \citealt{gutierrez17b}). Finally, the ejecta will become transparent (nebular phase), allowing the photons to escape from the core. Therefore, the spectrum will be dominated by emission lines formed by recombination or by collisional excitation such as \ion{O}{I} $\lambda\lambda 6300$, 6364, \ion{Fe}{II} $\lambda 7155$, and \ion{Ca}{II} $\lambda\lambda 7291$, 7323.

\subsubsection{Sample properties}\label{data_spec_red}

\begin{figure}
\includegraphics[width=1.0\columnwidth]{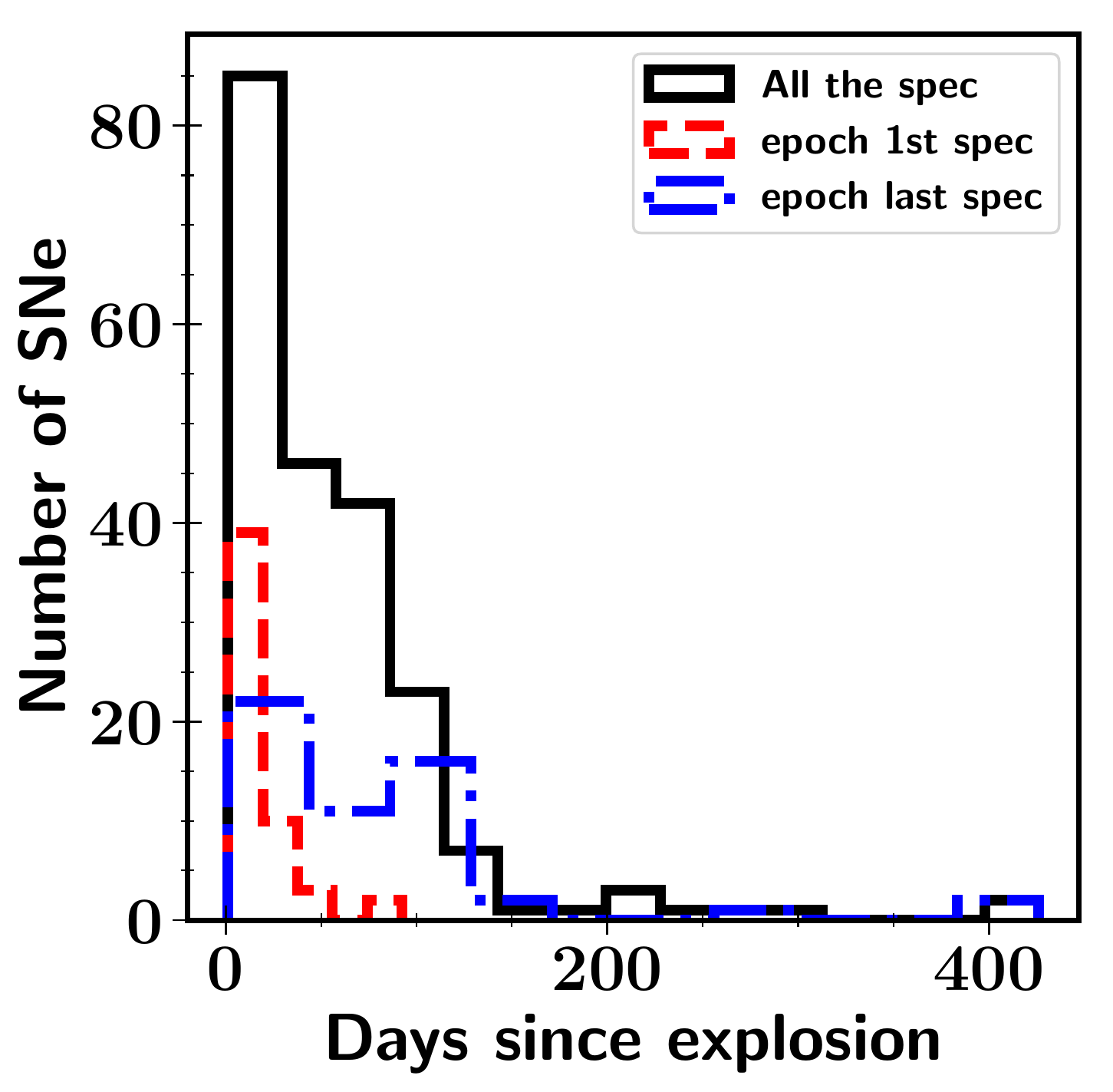}
\caption{Black histogram represents the spectrum epoch distributions. Red-dot histogram and blue dash-dot histogram are the distributions in days of the first and the last spectrum taken for each SN, respectively.}
\label{fig:data_spec_red}
\end{figure}

In Figure \ref{fig:spec_KAIT}, we display 213 optical spectra of 55 SNe~II, among which $\sim 160$ spectra from 43 SNe~II are previously unpublished. The distribution of our spectroscopic sample as a function of the epoch after the explosion is displayed in Figure \ref{fig:data_spec_red}, showing that the majority (87\%) of the spectra were taken $< 100$ days since the explosion and only 27 spectra were obtained after 100 days. Half of the spectra were obtained during the hydrogen recombination phase, between 30 and 100 days. The earliest spectrum in our compilation corresponds to SN~2013ft (iPTF13dkk) at $1 \pm 1$ days (already published by \citealt{khazov16}), followed by the unpublished SN~2016fqr spectrum at $2 \pm 1.5$ days. The oldest spectrum was taken at 4$26 \pm 19$ days (SN~2015C). In the same figure, we also represent for the first and last spectrum epoch (for each SN) distributions in red and blue, respectively. The majority of SNe in our compilation have their first spectra within 20 days after the explosion, with an average value of 18 days and a standard deviation of 17 days. The last spectrum was obtained on average 77 days after the explosion, and three unpublished SNe had their last spectra taken after 200 days (SN~2015C, SN~2015V, and SN~2015W), during the nebular phase.

\subsubsection{Median spectra}

In this section, following the work done by \citet{liu16} and \citet{shivvers19}, we construct a median spectrum at different phases to investigate the spectral line variations between the different SNe~II. As epochs, we choose 15, 50, 80, and $> 250$ days after the explosion, corresponding to the maximum brightness, the recombination phase, the end of the plateau, and the radioactive phase, respectively. For each epoch (at $\pm 5$ days except for the radioactive phase), we select only one spectrum per SN and then correct the spectrum for the MW extinction and the redshift. 

\begin{figure*}
\includegraphics[width=1.0\columnwidth]{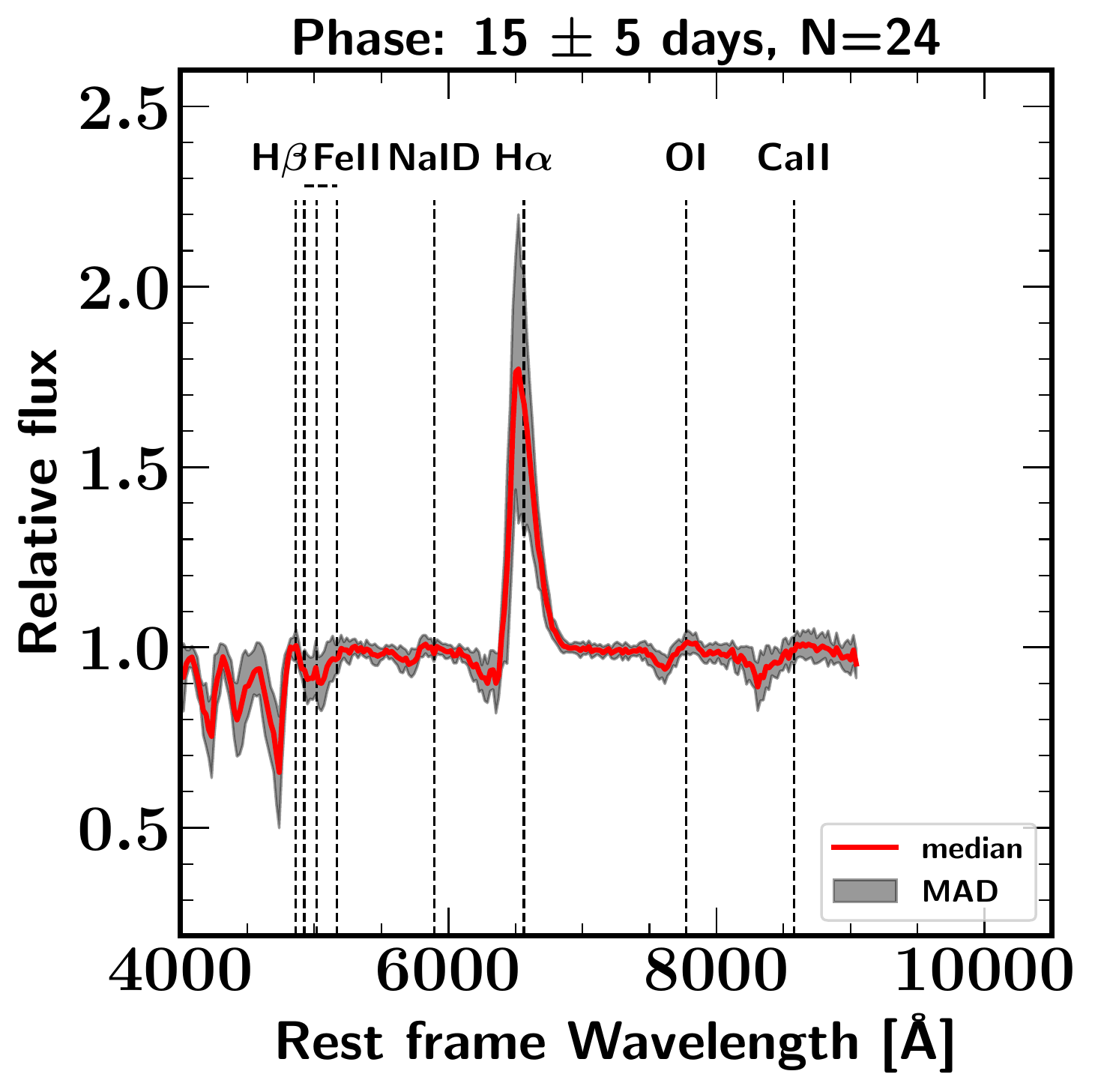}\includegraphics[width=1.0\columnwidth]{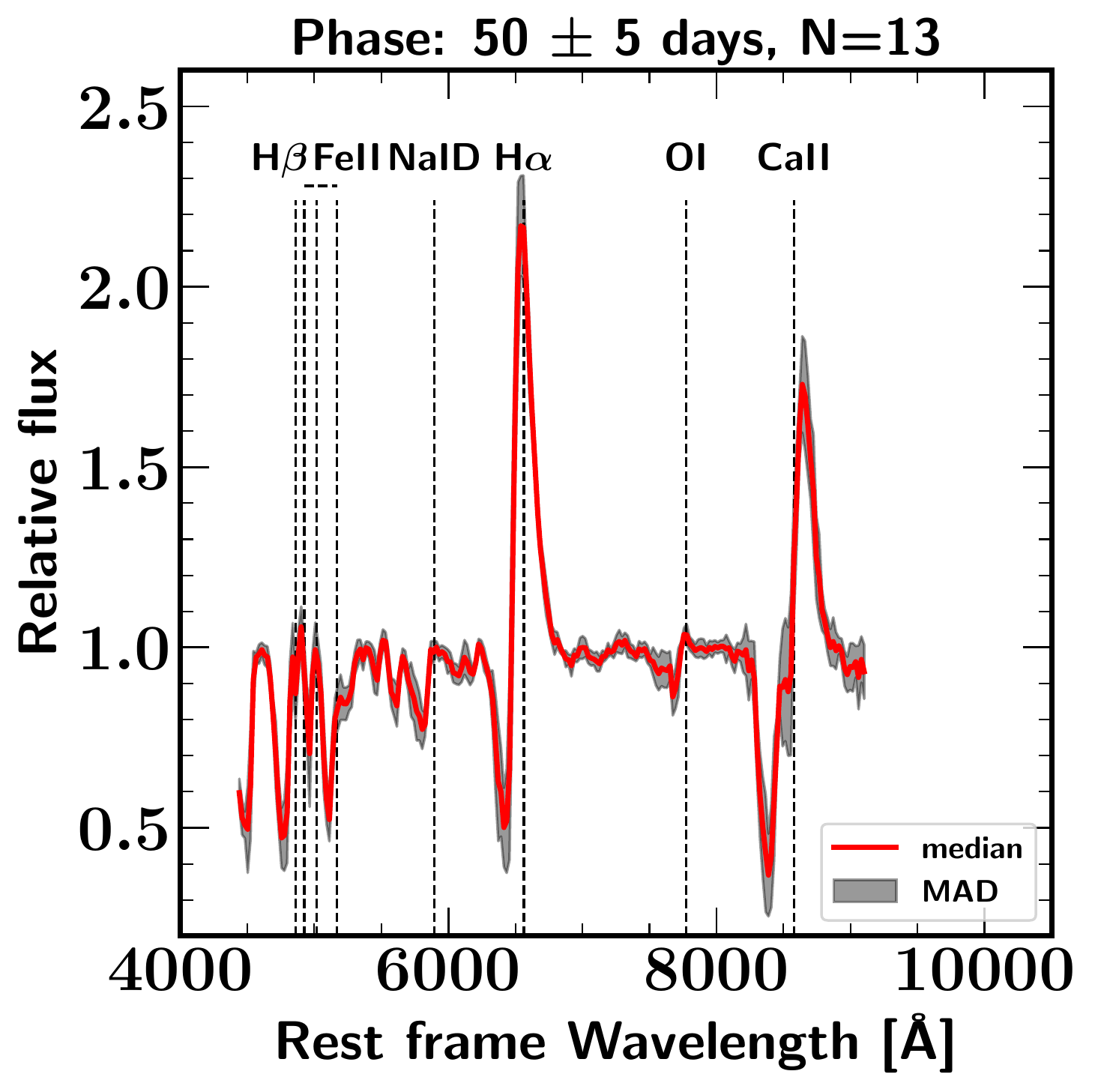}\\
\includegraphics[width=1.0\columnwidth]{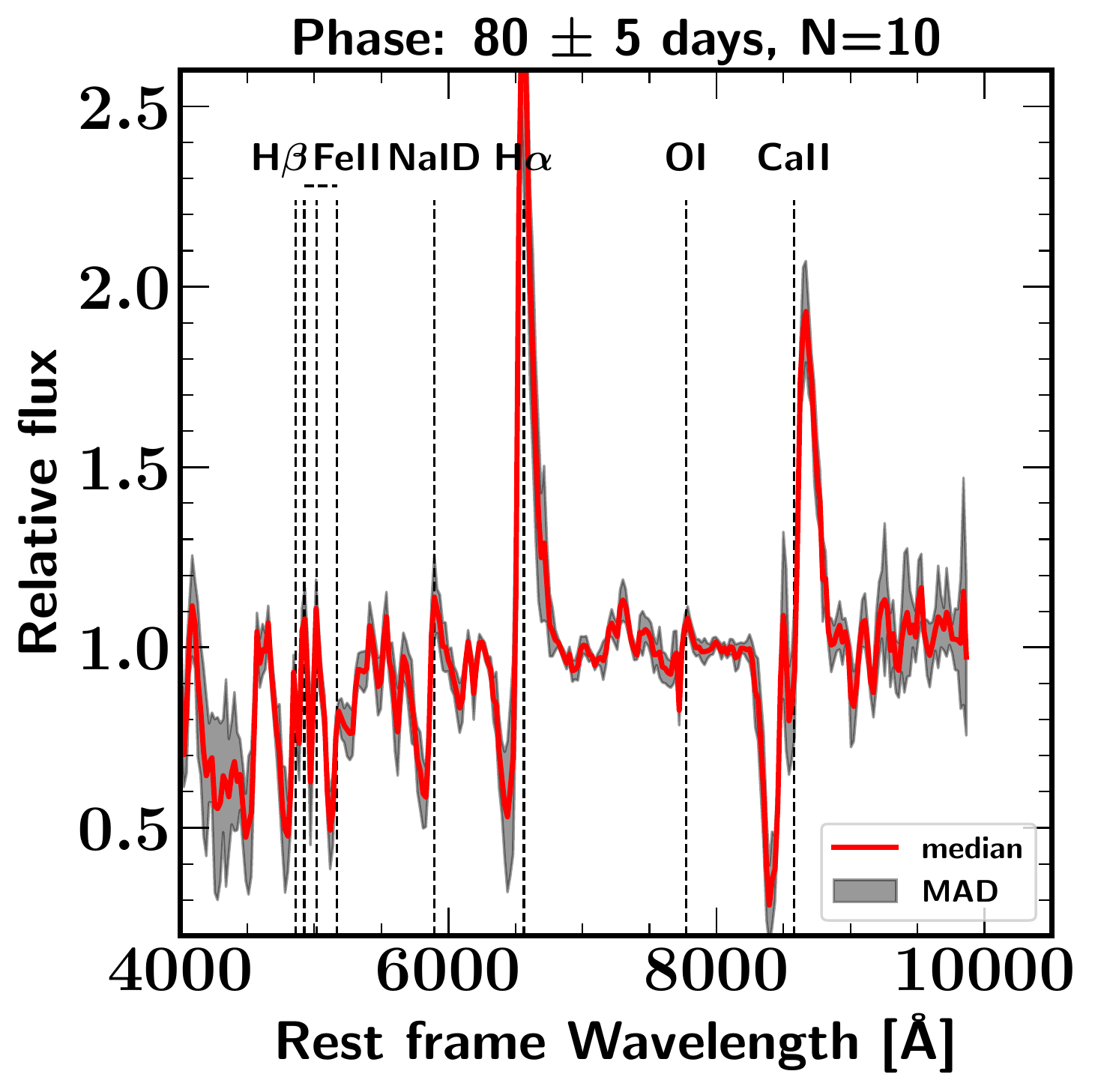}\includegraphics[width=1.0\columnwidth]{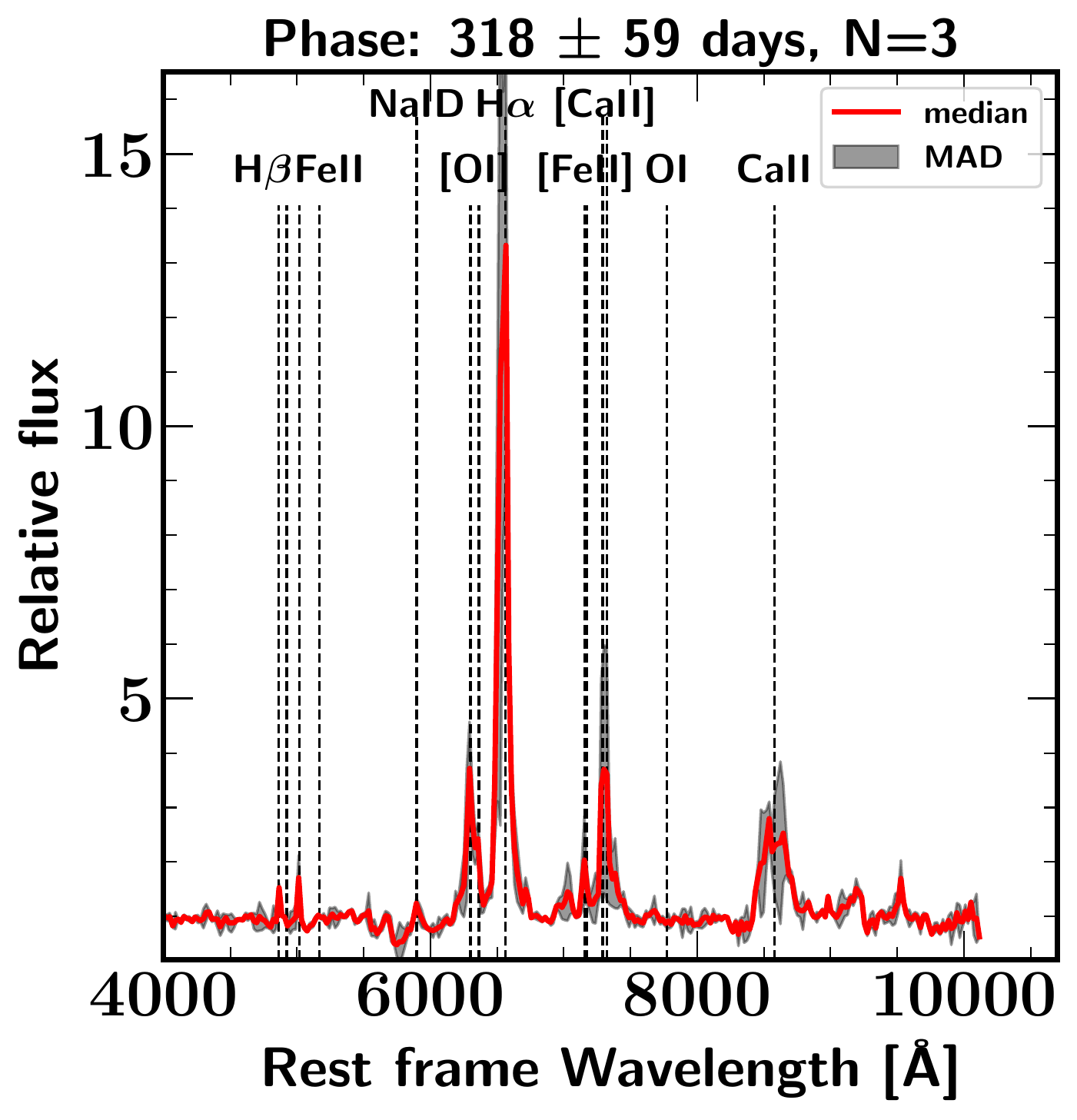}
\caption{Median spectra at 15 $\pm$ 5, 50 $\pm$ 5, 80 $\pm$ 5, and $\sim 300$ days after the explosion are displayed. Only one spectrum per SN is included, and the number of spectra used is indicated ($N$). The median spectrum is shown in red while the median absolute deviation is in grey.}
\label{fig:spec_median}
\end{figure*}

Each corrected spectrum is normalised using a pseudocontinuum defined with a cubic spline. Then, all the normalised spectra are smoothed (window of width 21~\AA) and then interpolate using the same wavelength array. Finally, for each wavelength, we derive the median flux value and its median absolute deviation. In Figure \ref{fig:spec_median}, the median spectra at the four different epochs are displayed.

During the photospheric phase (15, 50, 80 days), most of the spectral variation is seen in the H$\alpha$ line profile and the strength of iron lines. These variations reflect the diversity of SN~II progenitor properties. For example, the ratio between the absorption and the emission of the H$\alpha$ P-Cygni profile correlates with the expansion velocity \citep{gutierrez14}. Similarly, \citet{dessart14} has shown that metal line shapes depend on the progenitor metallicity. Note also, even if some variations are seen in the median spectra during the plateau phase (50 days), almost no differences are seen between the median spectra of the slow-declining and fast-declining SNe (7/13 SNe~II with $s_{2} > 1.5$ mag (100 days)$^{-1}$). This is again consistent with the fact that the SNe~II compose a unique group \citep{anderson14a,sanders15,valenti16,galbany16a}. Finally, in the nebular spectrum, the shapes of the different emission-line profiles are similar \citep{maguire12a}, but a variation of the strength of the forbidden emission lines (e.g., [\ion{O}{I}] $\lambda\lambda 6300$, 6364) is seen. This scatter indicates differences in progenitor-star masses.

\subsubsection{Absorption velocity}\label{txt:vel}

In this and the following (\ref{txt:pEW}) sections, we measure the absorption velocity and the strength of six spectral lines present during the photospheric phase. Two lines are visible throughout the whole SN spectrum evolution (H$\alpha$ $\lambda 6563$, H$\beta$ $\lambda 4861$), and four during the plateau phase (\ion{Fe}{II} $\lambda\lambda 4924$, 5018, 5169 and \ion{Na}{I}~D $\lambda 5893$). We also investigate the H$\alpha$ extra absorption component (also called ``Cachito''; \citealt{gutierrez17b}) which is related to \ion{Si}{II} $\lambda 6355$ at early phases and to a high-velocity feature of H$\alpha$ at late epochs \citep{gutierrez17b}. Even if after 20 days, strong lines such as \ion{O}{I} $\lambda 7774$ or the \ion{Ca}{II} near-infrared triplet ($\lambda\lambda 8498$, 8542, 8662) emerge in the spectrum, those lines are not included in this analysis as they are contaminated by telluric lines or result from blended lines, making it difficult to measure the pseudoequivalent width (pEW) and the velocity.

Minimum flux of the absorption of different features is used to measure the ejecta expansion velocities. The minimum flux position in wavelength is estimated using IRAF and by fitting a Gaussian profile. The position in wavelength is then transformed into velocity using the Doppler relativistic formula. To obtain velocity uncertainties, we change the continuum fit many times, measure the minimum of the absorption and determining their standard deviation. To this uncertainty, another one from the spectral resolution ($\sim 10$ \AA) is also added in quadrature. All of the velocities are shown in Figure \ref{fig:velocity_evolution} together with their average evolution from the Carnegie Supernova Project-I (CSP-I) SN~II sample \citep{gutierrez17b}.

Figure \ref{fig:velocity_evolution} shows that SN~II ejecta velocities follow the typical evolution for homologous expansion (a power law; \citealt{hamuyphd}): in the ejecta, deeper material is at smaller radii and therefore moving at lower velocities. At all epochs, H$\alpha$ shows higher velocities than other lines, with a velocity starting at $\sim 10,500$ km s$^{-1}$ at early times (10 days) to $\sim 6500$ km s$^{-1}$ during the plateau phase (50 days). At 50 days, the H$\alpha$ velocity displays a large range from $\sim 8500$ km s$^{-1}$ to $\sim 4500$ km s$^{-1}$. Following H$\alpha$, H$\beta$ has the highest velocities, with a velocity ranging from $\sim 8500$ km s$^{-1}$ (10 days) to $\sim 5500$ km s$^{-1}$, on average 1000--1500 km s$^{-1}$ smaller than H$\alpha$. Finally, the iron lines exhibit the lowest velocities, with a range from $\sim 6000$ km s$^{-1}$ to $\sim 2500$ km s$^{-1}$ at 50 days. This velocity sequence (H$\alpha$ $>$ H$\beta$ $>$ \ion{Fe}{ii}) is expected; since the H$\alpha$ and H$\beta$ lines are formed at larger radii, and therefore their velocities should be higher than those formed closer to the photosphere like the \ion{Fe}{ii} lines \citep{dessart05,takats12}. As seen in Figure \ref{fig:velocity_evolution}, our velocity measurements are consistent with those derived by \citet{gutierrez17b} using a low-redshift sample of 122 SNe~II. For each element, the majority of our velocities are within their $1\sigma$ standard deviation (red filled region), and their average values are also similar to ours (e.g., median difference of $\sim 200$ km s$^{-1}$ for H$\alpha$).

It is also important to note that as suggested by a number of previous studies \citep{nugent06,poznanski10,takats12,dejaeger17a}, H$\beta$ absorption is the best line for measuring the expansion velocity of the ejecta at early times or in a noisy spectrum. At early times, H$\alpha$ absorption is sometimes blended with \ion{Si}{II} $\lambda 6355$ \citep{gutierrez17b}, leading to an overestimate of the velocities ($> 14,000$ km s$^{-1}$; see Figure \ref{fig:velocity_evolution}); moreover, the \ion{Fe}{II} $\lambda$5018 line only appears later than H$\beta$ (30--40 days after the explosion), while \ion{Fe}{II} $\lambda$5169 is often blended with other features.

Finally, as discussed by \citep{gutierrez17b}, 40\% of the SNe~II in our compilation exhibit on the blue side of the H$\alpha$ lines an extra component between 6100 and 6300~\AA\ at early epochs ($< 40$--45 days after explosion) and between 6300 and 6450 \AA\ at later epochs. The differences in the line shape and strength between the two phases suggest different origins: at early epochs, the extra component is associated with \ion{Si}{II} $\lambda 6355$, while at later epochs it is associated with a high-velocity feature of H$\alpha$ \citep{gutierrez17b}.

\subsubsection{Absorption-line strength measurements}\label{txt:pEW}

Absorption-line strength measurements are useful for a better understanding of SN~II progenitor diversity: metal-line strength depends on progenitor metallicity \citep{dessart14} and plays a role in the Hubble-diagram scatter \citep{dejaeger17b}. To quantify the absorption strength, we use the pEW. As for the velocities, the pEW is derived using IRAF by marking the two edges of the absorption line and defining a pseudocontinuum. Then a pixel-value integration is achieved between the two marked points. 

\begin{figure*}
\centering
\includegraphics[width=0.99\textwidth]{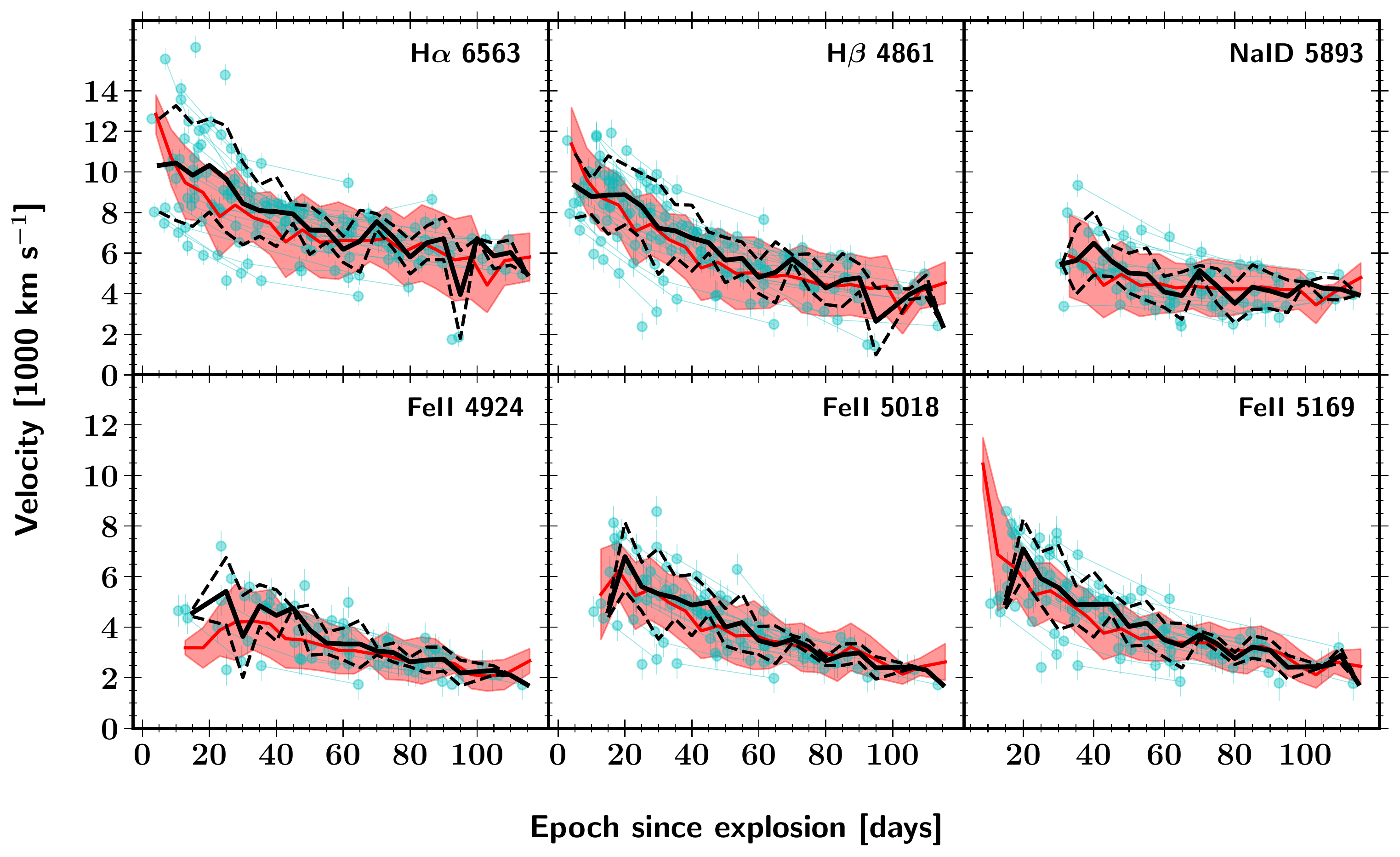}
\caption{Expansion-velocity evolution of H$\alpha$ $\lambda 6563$, H$\beta$ $\lambda 4861$, \ion{Na}{I}~D $\lambda 5893$, and \ion{Fe}{II} $\lambda\lambda 4924$, 5018, 5169 are displayed (cyan circles). The black solid and dashed lines represent the average velocity in bins of five days and its standard deviation. The solid red and filled region are the average and the standard deviation derived by \citet{gutierrez17b} using the CSP-I sample.}
\label{fig:velocity_evolution}

\includegraphics[width=0.99\textwidth]{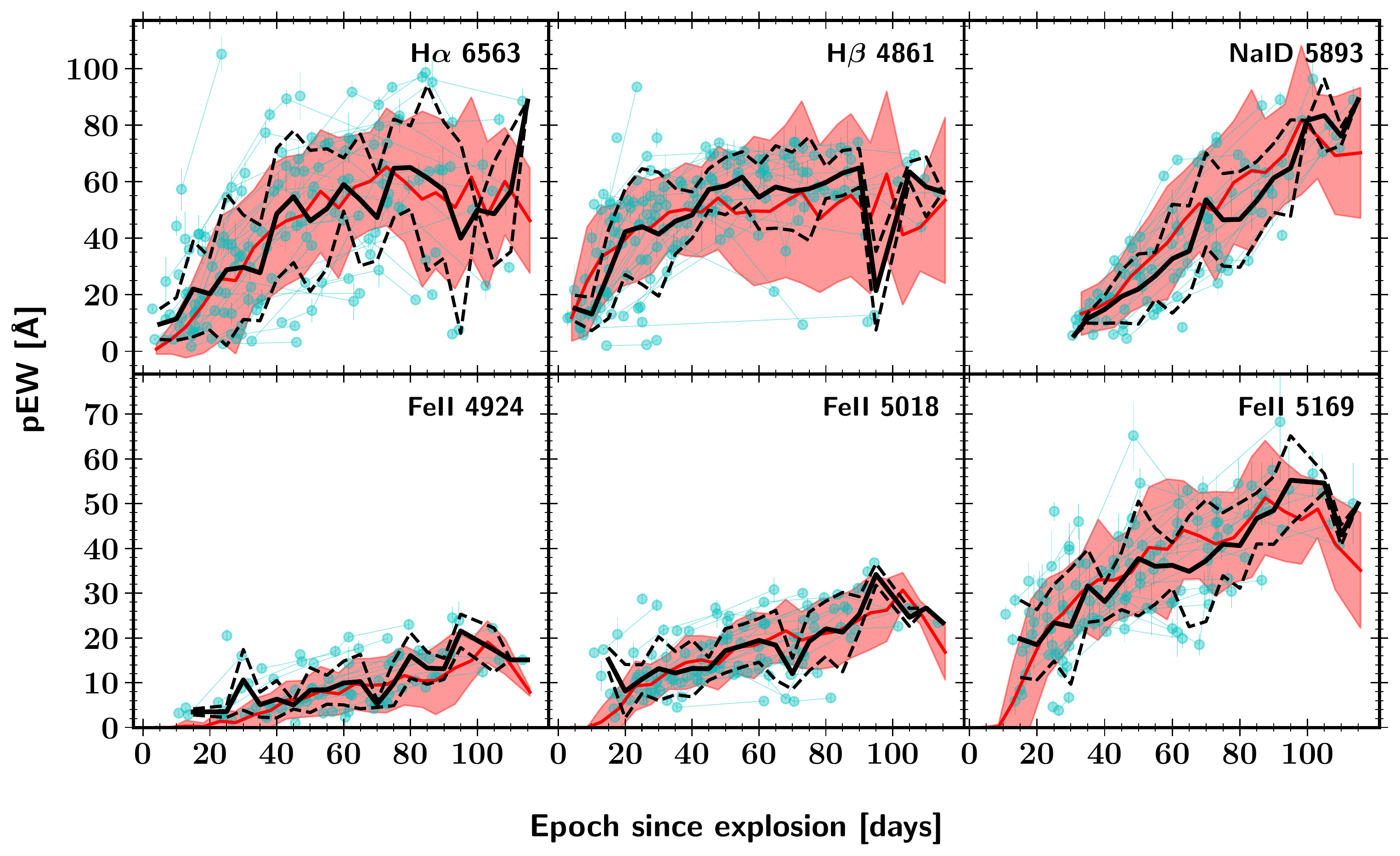}
\caption{Pseudoequivalent-width evolution of H$\alpha$ $\lambda 6563$, H$\beta$ $\lambda 4861$, \ion{Na}{I}~D $\lambda 5893$, and \ion{Fe}{II} $\lambda\lambda 4924$, 5018, 5169 are displayed (cyan circles). The black solid and dashed lines represent the average pEW in bins of five days and its standard deviation. The solid red and filled region are the average and the standard deviation derived by \citet{gutierrez17b} using the CSP-I sample.}
\label{fig:EW_evolution}
\end{figure*}

In Figure \ref{fig:EW_evolution}, the pEW evolution for each element is displayed together with the average evolution from the CSP-I SN~II sample \citep[][shown in red]{gutierrez17b}. H$\alpha$ and H$\beta$ exhibit similar evolutionary behaviour, with an increment of the pEW during the first two months from 0 to $\sim 80$ \AA\ following by a plateau. However, as noted by \citep{gutierrez17b}, after $\sim 80$ days the peW decreases in a few SNe~II. On the other hand, the \ion{Fe}{II} and \ion{Na}{I}~D pEW evolution show a steady increase with time. For \ion{Fe}{II} $\lambda$5169, the pEW grows from 0 to $\sim 60$ \AA, becoming the strongest iron line, while the \ion{Na}{I}~D $\lambda 5893$ pEW evolves from from 0 to $\sim 90$ \AA.

\subsubsection{Late-time spectra}

After being powered by hydrogen recombination, the light curve enters a phase where the hydrogen envelope becomes transparent and the core becomes visible. At that time, the energy is produced by the radioactive decay of $^{56}$Co into $^{56}$Fe. The P-Cygni absorption features (indicators of an optically thick photosphere) present in the spectrum disappear, leaving a weak continuum dominated by forbidden emission lines of oxygen ([\ion{O}{I}] $\lambda\lambda 6300$, 6364), calcium ([\ion{Ca}{II}] $\lambda\lambda 7291$, 7325), iron ([\ion{Fe}{II}] $\lambda 7155$), and lines that were present during the photospheric phase such as H$\alpha$ $\lambda 6563$ and the \ion{Ca}{II} near-infrared triplet ($\lambda\lambda 8498$, 8542, 8662). Nebular spectra are useful for constraining the physical properties of the SN progenitor. After carefully taking into account the primordial oxygen, the [\ion{O}{I}] $\lambda\lambda 6300$, 6364 flux can be used to estimate the main-sequence mass of the progenitor star; larger progenitor masses lead to stronger oxygen lines \citep{maguire12a,jerkstrand12,jerkstrand14,dessart19a}. 

In this section, we compare our late-time spectra to a set of theoretical nebular spectra presented by \citet{jerkstrand14} and \citet{dessart13}. For the first set, four progenitor masses have been modelled (12, 15, 19, and 25 ${\rm M}_{\odot}$), while for the six models presented by \citet{dessart13} the mass is constant (15 ${\rm M}_{\odot}$) but the progenitor metallicity and mixing-length parameters vary.

Most of the nebular spectra from our sample have already been published by \citet{silverman17}. However, seven spectra of three recent SNe~II were previously unpublished (SN~2015C, SN~2015V, and SN~2015W). These spectra were selected based on their epochs ($>200$ days after the explosion) and visual inspection (no continuum emission or P-Cygni absorption features). We compare each spectrum to each model at the closest epoch, and select the best fit using a cross-correlation algorithm over the full wavelength range and a visual sanity check.

\begin{figure*}
\includegraphics[width=1.0\textwidth]{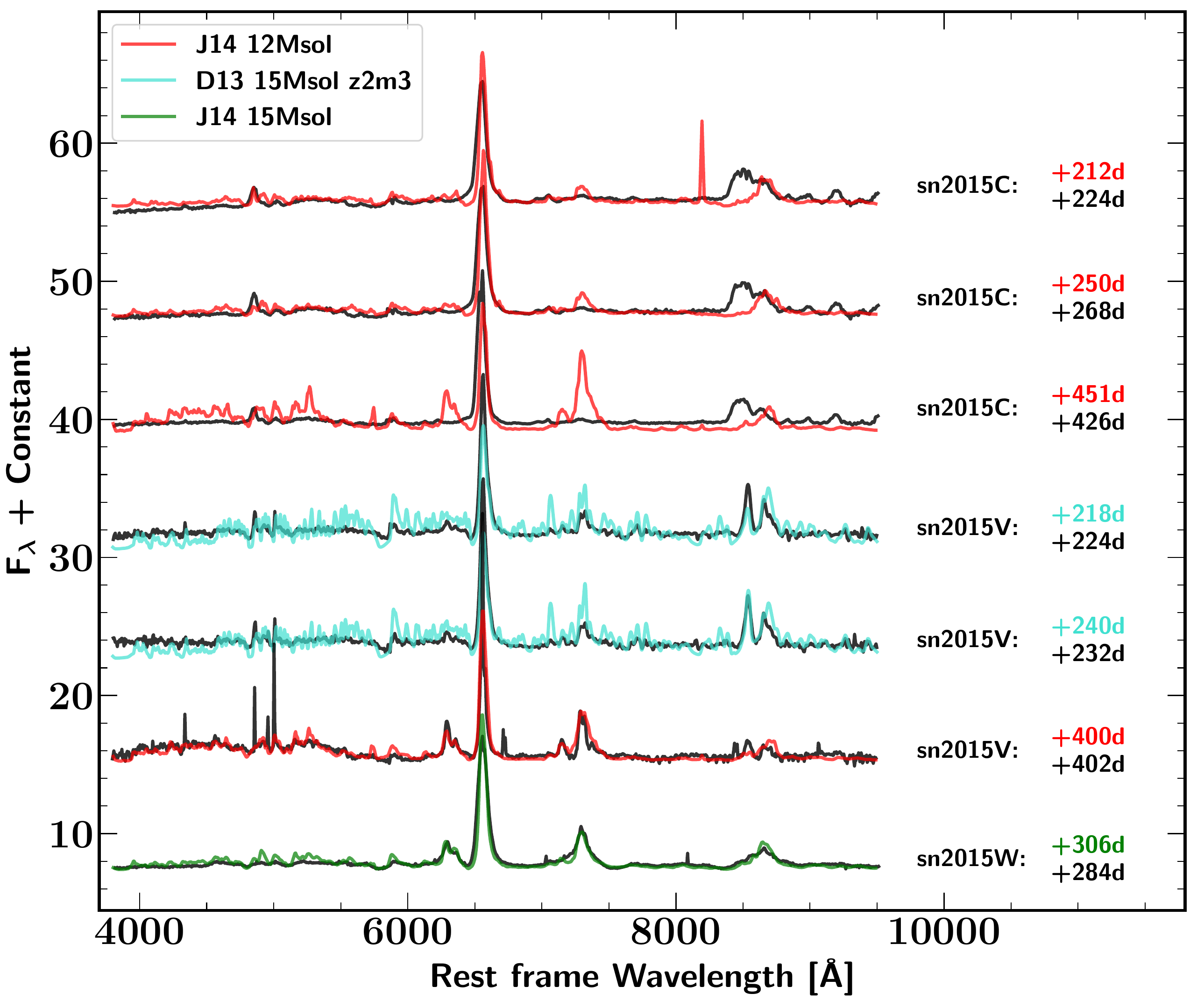}
\caption{Eight observed nebular spectra of four SNe~II from our sample (in black) are compared with theoretical models from \citet{dessart13} or \citet{jerkstrand14}. The epochs of the observed spectra and the best theoretical models are shown together with the SN names.}
\label{fig:nebular_spec}
\end{figure*}

In Figure \ref{fig:nebular_spec}, the seven nebular spectra together with their best theoretical fits are displayed. Consistent with \citet{jerkstrand15} and archival prediscovery images \citep{smartt15}, our nebular spectra are in good agreement with progenitor stars having $M < 16 {\rm M}_{\odot}$. Even if (as noticed by \citealt{silverman17}) theoretical models generally underproduce the \ion{Ca}{II} NIR triplet ($\lambda\lambda 8498$, 8542, 8662) or overproduce the \ion{He}{I} ($\lambda 7065$) emission, the strength of the [\ion{O}{I}] $\lambda\lambda 6300$, 6364 doublet is well fitted by the 12 or 15 ${\rm M}_{\odot}$ models. It is worth noting the case of SN~2015C, where almost no spectral evolution is seen between the spectra taken 268 and 426 days after the explosion, while theoretical models show strong evolution of the [\ion{Ca}{II}] $\lambda\lambda 7291$, 7324 flux.

\section{Summary and Conclusions}\label{txt:conclusions}

In this work, we present a compilation of SNe~II observed over the past decade by the Berkeley SN group. This sample consists of 55 optical light curves obtained with the KAIT and Nickel telescopes at Lick Observatory. Our $BVRI$ light curves start (on average) 12
days after the explosion and last until 144 days. For each band, we estimate the main photometric parameters (the absolute peak magnitude, the length of the plateau, and the slope of the plateau), and we also investigate the $(B-R)$, $(B-I)$, $(B-V)$, $(V-I)$, $(V-R)$, and $(R-I)$ colour curves. In addition to the visual-wavelength photometry, 213 spectra ranging between 1 and 426 days post-explosion are analysed. For each spectrum, we measure the absorption velocity and the strength of six spectral lines: H$\alpha$ $\lambda 6563$, H$\beta$ $\lambda 4861$, \ion{Fe}{II} $\lambda\lambda 4924$, 5018, 5169, and \ion{Na}{I}~D $\lambda 5893$. To study the spectral-line variations among the different SNe~II, we also construct a median spectrum at four different phases (15, 50, 80, and older than 250 days after the explosion). Finally, we compare our seven previously unpublished late-time spectra to a set of theoretical nebular spectra. The main results obtained from our photometric and spectroscopic analysis can be summarised as follows.

\begin{enumerate}
\item{Confirming earlier studies, we find that SNe~IIP and SNe~IIL share common photometric and spectroscopic properties and therefore, form a continuous group.} 

\item{The absolute peak magnitudes (not corrected for host-galaxy extinction) found are $<B_{\rm max}>=-16.40$ mag ($\sigma=1.08$, $N=42$), $<V_{\rm max}>=-16.54$ mag ($\sigma=0.95$, $N=42$), $<R_{\rm max}>=-16.78$ mag ($\sigma=0.90$, $N=41$), and $<I_{\rm max}>=-16.97$ mag ($\sigma=0.90$, $N=42$).}

\item{Similar to previous studies \citep{faran14a,galbany16a,dejaeger18a}, we found that redder colours (e.g., $R-I$) increase more slowly with time than the bluer colours (e.g., $B-V$) as the blue part of the spectrum is more sensitive to temperature changes. At a given epoch, the scatter among different SNe is larger for bluer colours than redder colours (cf Figure \ref{fig:colour_evolution}). This scatter could be caused by intrinsic progenitor properties or host-galaxy extinction.}

\item{The plateau length is similar in all the bands while the plateau slope decreases in redder filters \citep{sanders15,galbany16a}.}

\item{For each band, the plateau slope correlates with the plateau length and the absolute peak magnitude: SNe~II with steeper decline are generally brighter and have shorter plateau duration \citep{anderson14a,sanders15,galbany16a}.}

\item{In SN~II photospheric spectra, most of the variation is seen in the H$\alpha$ feature and the strength of iron lines, reflecting the diversity of SN~II progenitor properties (e.g., \citealt{gutierrez14,dessart14}).}

\item{Consistent with \citep{gutierrez17b}, the pEW of H$\alpha$ and H$\beta$ increases during the first two months from 0 to $\sim 80$ \AA\ until reaching a plateau, while the \ion{Fe}{II} and \ion{Na}{I}~D pEW show a steady increase with time.}

\item{Our nebular spectra are in good agreement with progenitor stars having $M <16 {\rm M}_{\odot}$.}

\end{enumerate}

Note that, our photometry and spectroscopy is available for download from the Berkeley SuperNova DataBase (SNDB\footnote{\url{http://heracles.astro.berkeley.edu/sndb/}}; \citealt{silverman12}) or can be requested from the authors. All of the spectra are in units of $10^{-15}$ erg s$^{-1}$ cm$^{-2}$ \AA$^{-1}$. The photometry is published in the natural system. The photometric error bars include only the statistical uncertainties (scatter in sky values, Poisson errors) and uncertainties in the calibration catalogue. No uncertainties associated with the host-galaxy subtraction are applied ($\sim 0.06$ mag; see Stahl et al. 2019, in prep.). 

\section*{Acknowledgements}

The anonymous referee is thanked for their thorough reading of the manuscript, which helped clarify and improve the paper. We are grateful to the staff at Lick Observatory for their expert assistance. KAIT and its ongoing operation were made possible by donations from Sun Microsystems, Inc., the Hewlett-Packard Company, Auto Scope Corporation, Lick Observatory, the National Science Foundation (NSF), the University of California, the Sylvia \& Jim Katzman Foundation, and the TABASGO Foundation. A major upgrade of the Kast spectrograph on the Shane 3~m telescope at Lick Observatory was made possible through generous gifts from the Heising-Simons Foundation as well as William and Marina Kast. 

Research at Lick Observatory is partially supported by a generous gift from Google. Support for A.V.F.'s supernova group has also been provided by the NSF, Marc J. Staley (B.E.S. is a Marc J. Staley Graduate Fellow), the Richard and Rhoda Goldman Fund, the TABASGO Foundation, Gary and Cynthia Bengier (T.deJ. is a Bengier Postdoctoral Fellow), the Christopher R. Redlich Fund, and the Miller Institute for Basic Research in Science (U.C. Berkeley). In addition, we greatly appreciate contributions from numerous individuals, including
Charles Baxter and Jinee Tao,
George and Sharon Bensch 
Firmin Berta,     
Marc and Cristina Bensadoun, 
Frank and Roberta Bliss,   
Eliza Brown and Hal Candee,
Kathy Burck and Gilbert Montoya,
Alan and Jane Chew,        
David and Linda Cornfield,
Michael Danylchuk,        
Jim and Hildy DeFrisco,   
William and Phyllis Draper,
Luke Ellis and Laura Sawczuk,
Jim Erbs and Shan Atkins,    
Alan Eustace and Kathy Kwan, 
Peter and Robin Frazier 
David Friedberg,             
Harvey Glasser,              
Charles and Gretchen Gooding,
Alan Gould and Diane Tokugawa,
Thomas and Dana Grogan,
Timothy and Judi Hachman 
Alan and Gladys Hoefer,
Charles and Patricia Hunt, 
Stephen and Catherine Imbler,
Adam and Rita Kablanian, 
Roger and Jody Lawler,
Kenneth and Gloria Levy,     
Peter Maier, 
DuBose and Nancy Montgomery,
Rand Morimoto and Ana Henderson,
Sunil Nagaraj and Mary Katherine Stimmler, 
Peter and Kristan Norvig,   
James and Marie O'Brient,  
Emilie and Doug Ogden,   
Paul and Sandra Otellini,     
Jeanne and Sanford Robertson,
Sissy Sailors and Red Conger 
Stanley and Miriam Schiffman,
Thomas and Alison Schneider,
Ajay Shah and Lata Krishnan, 
Alex and Irina Shubat,     
the Silicon Valley Community Foundation,
Mary-Lou Smulders and Nicholas Hodson,
Hans Spiller,
Alan and Janet Stanford,
the Hugh Stuart Center Charitable Trust,
Clark and Sharon Winslow,
Weldon and Ruth Wood,
David and Angie Yancey, 
and many others.
M.L.G. acknowledges support from the DIRAC Institute in the Department of Astronomy at the University of Washington. The DIRAC Institute is supported through generous gifts from the Charles and Lisa Simonyi Fund for Arts and Sciences, and the Washington Research Foundation. M.M. is supported by NSF CAREER award AST--1352405, by NSF award AST--1413260, and by a Humboldt Faculty Fellowship. X.W. is supported by the National Natural Science Foundation of China (NSFC grants 11325313, 11633002, and 11761141001), and the National Program on Key Research and Development Project (grant 2016YFA0400803). X.G. is supported by the National Natural Science Foundation of China (NSFC grant 11673006) and the Guangxi Science Foundation (grants 2016GXNSFFA380006 and 2017AD22006).

We thank the following people (mostly U.C. Berkeley undergraduate students) for their effort in taking Lick/Nickel and Lick/Shane data: 
Carmen Anderson,
Iair Arcavi, 
Aaron Barth, 
Misty Bentz, 
Joshua Bloom, 
Azalee Bostroem, 
Stanley Brown, 
Jieun Choi, 
Nick Choski, 
Michael Cooper, 
Louis Desroches, 
Josh Emery, 
Matt George, 
Christopher Griffith,
Jenifer Gross,
Andrew Halle,
Romain Hardy, 
Deam Hiner, 
Anthony Khodanian, 
Robert Kibrick, 
Michelle Kislak, 
Io Kleiser, 
Jason Kong, 
Daniel Krishnan,
Laura Langland Shula, 
Joel Leja, 
Brent Macomber,
Adam Miller, 
Shaunak Modak, 
Robin Mostardi, 
Yukei Murakami,
Andrew Rickter, 
Frank Serduke, 
Josh Shiode, 
Brian Siana, 
Jackson Sipple, 
Diamond-Stanic, 
Thea Steele,
Haynes Stephens,
Stephen Taylor, 
Patrick Thrasher, 
Brad Tucker, 
Vivian U, 
Stefano Valenti, 
Jonelle Walsh, 
Jeremy Wayland, 
Dustin Winslow,
Diane Wong, 
Jong-Hak Woo, and 
Yinan Zhu.

This research has made use of the NASA/IPAC Extragalactic Database (NED), which is operated by the Jet Propulsion Laboratory, California Institute of Technology, under contract with the National Aeronautics and Space Administration (NASA). The Pan-STARRS1 Surveys (PS1) and the PS1 public science archive have been made possible through contributions by the Institute for Astronomy, the University of Hawaii, the Pan-STARRS Project Office, the Max-Planck Society and its participating institutes, the Max Planck Institute for Astronomy, Heidelberg and the Max Planck Institute for Extraterrestrial Physics, Garching, The Johns Hopkins University, Durham University, the University of Edinburgh, the Queen's University Belfast, the Harvard-Smithsonian Center for Astrophysics, the Las Cumbres Observatory Global Telescope Network Incorporated, the National Central University of Taiwan, the Space Telescope Science Institute, NASA under grant  NNX08AR22G issued through the Planetary Science Division of the NASA Science Mission Directorate, NSF grant AST--1238877, the University of Maryland, Eotvos Lorand University (ELTE), the Los Alamos National Laboratory, and the Gordon and Betty Moore Foundation. 





\appendix
\section{SN~II sample information, light curves, and spectra.}\label{AppendixA}
\onecolumn
\setlength\LTleft{-1.2cm}
\scriptsize
\begin{longtable}{lcccccccccl}
\caption{Type II Supernova Sample.}\\
\hline
\hline 
\multicolumn{1}{l}{SN} & \multicolumn{1}{c}{Host Galaxy} & \multicolumn{1}{c}{$A_{V}$(MW)} & \multicolumn{1}{c}{$v_{\rm helio}$}& \multicolumn{1}{c}{DM} & \multicolumn{1}{c}{Explosion date} & \multicolumn{1}{c}{Nondetect} & \multicolumn{1}{c}{Discovery} & \multicolumn{1}{c}{$\#$ Phot} & \multicolumn{1}{c}{$\#$ Spectra} & \multicolumn{1}{l}{Ref}\\
\multicolumn{1}{l}{} & \multicolumn{1}{c}{} & \multicolumn{1}{c}{mag} & \multicolumn{1}{c}{km s$^{-1}$} & \multicolumn{1}{c}{mag} & \multicolumn{1}{c}{MJD} & \multicolumn{1}{c}{UT} & \multicolumn{1}{c}{UT} & \multicolumn{1}{c}{} & \multicolumn{1}{c}{} & \multicolumn{1}{l}{}\\
\hline
\endfirsthead
\hline
\hline 
\multicolumn{1}{l}{SN} & \multicolumn{1}{c}{Host Galaxy} & \multicolumn{1}{c}{$A_{V}$(MW)} & \multicolumn{1}{c}{$v_{\rm helio}$} & \multicolumn{1}{c}{DM}& \multicolumn{1}{c}{Explosion date} & \multicolumn{1}{c}{Nondetect} & \multicolumn{1}{c}{Discovery} & \multicolumn{1}{c}{$\#$ Phot} & \multicolumn{1}{c}{$\#$ Spectra} & \multicolumn{1}{l}{Ref}\\
\multicolumn{1}{l}{} & \multicolumn{1}{c}{} & \multicolumn{1}{c}{mag} & \multicolumn{1}{c}{km s$^{-1}$} & \multicolumn{1}{c}{mag} & \multicolumn{1}{c}{MJD} & \multicolumn{1}{c}{UT} & \multicolumn{1}{c}{UT} & \multicolumn{1}{c}{} & \multicolumn{1}{c}{} & \multicolumn{1}{l}{}\\
\hline
\endhead
SN 2006ee & NGC 774 &0.167 &4620 &34.12(0.14) &53961.0(4) &Aug. 10.45 &Aug. 18.47 &30 &1 &\citet{cbet-06ee} \\
SN 2006ek & MCG +04-52-3 &0.213 &6104 &34.61(0.11) &53968.5(4) &Aug. 17.39 &Aug. 25.27 &12 &1 &\citet{cbet-06ek} \\
SN 2007ck & MCG +05-43-16 &0.309 &8083 &35.32(0.09) &54228.0(13)$^{a}$ &06 Sep. 18 &May 19.04 &122 &6 &\citet{cbet-07ck} \\
SN 2007il & IC 1704 &0.129 &6454 &34.75(0.11) &54348.5(4) &Sep 2.45 &Sep. 10.45 &16 &12$^{b}$ &\citet{cbet-07il} \\
SN 2007od & UGC 12846 &0.100 &1734 &31.91(0.80)$^{c}$ &54400.5(5)$^{a}$ &06 Oct 18 &Nov 2.85 &90 &3 &\citet{cbet-07od} \\
SN 2008aw & NGC 4939 &0.111 &3110 &33.48(0.18) &54517.5(10) &Feb. 11.54 &Mar. 2.49 &79 &5 &\citet{cbet-08aw} \\
SN 2008bx & Anon. &0.065 &2518 &32.99(0.80)$^{c}$ &54576.5(4) &Apr. 9 &Apr. 22.35 &106 &4 &\citet{cbet-08bx} \\
SN 2008ea & NGC 7624 &0.366 &4275 &33.77(0.16) &54646.5(8) &Jun. 21.44 &Jul. 6.07 &81 &5 &\citet{cbet-08ea} \\
SN 2008ex & UGC 11428 &0.201 &3945 &33.67(0.17) &54692.5(2) &Aug. 14.31 &Aug. 17.32 &70 &2 &\citet{cbet-08ex} \\
SN 2008gi & CGCG 415-004 &0.181 &7328 &35.07(0.09) &54742.5(9) &Sep. 24.40 &Oct. 12.41 &38 &2 &\citet{cbet-08be} \\
SN 2008if & MCG -01-24-10 &0.090 &3440 &33.68(0.16) &54806.3(5) &Dec. 2.23 &Dec. 12.21 &24 &1 &\citet{cbet-08if} \\
SN 2008in & NGC 4303 &0.061 &1566 &30.38(0.47)$^{c}$ &54824.5(2) &Dec. 23.95 &Dec. 26.79 &86 &3 &\citet{cbet-08in} \\
SN 2009N & NGC 4487 &0.057 &1036 &31.49(0.40)$^{c}$ &54846.5(5)$^{a}$ & Jan. 3 &Jan. 24.8 &18 &1 &\citet{cbet-09N} \\
SN 2009ao & NGC 2939 &0.106 &3339 &33.62(0.17) &54890.1(4) &Feb. 24.12 &Mar. 4.12 &67 &6$^{b}$ &\citet{cbet-09ao} \\
SN 2009at & NGC 5301 &0.047 &1503 &31.71(0.20)$^{c}$ &54900.5(8)$^{a}$ &05 Jan. 10 &Mar. 11.63 &51 &3 &\citet{cbet-09at} \\
SN 2009ay & NGC 6479 &0.117 &6650 &34.90(0.10) &54894.5(15) &Feb. 17 &Mar. 20.41 &58 &2 &\citet{cbet-09ay} \\
SN 2009hz & UGC 11499 &0.469 &7572 &35.16(0.08) &55043.8(3) &Jul 29.37 &Aug. 3.30 &42 &1 &\citet{cbet-09hz} \\
SN 2009js & NGC 918 &0.968 &1507 &32.33(0.15)$^{d}$ &55109.5(6) &Sep. 30.44 &Oct. 11.44 &105 &3 &\citet{cbet-09js} \\
SN 2009kr$^{e}$ & NGC 1832 &0.200 &1939 &32.08(0.25)$^{c}$ &55132.5(10)$^{a}$ &Oct. 3.76 &Nov. 6.73 &117 &7 &\citet{cbet-09kr}\\
SN 2010id$^{e}$ & NGC 7483 &0.166 &4939 &34.10(0.14) &55452.3(3)$^{f}$ &Sep. 11.34 &Sep. 15.24 &72 &1 &\citet{cbet-10id} \\
SN 2011cj & UGC 9356 &0.072 &2224 &32.86(0.47)$^{c}$ &55688.4(2) &May 5.39 &May 9.39 &106 &5 &\citet{cbet-11cj} \\
SN 2011ef & UGC 12640 &0.188 &4009 &33.60(0.18) &55759.5(1) &Jul. 16.44 &Jul. 18.46 &97 &3 &\citet{cbet-11ef} \\
SN 2011fd & NGC 2273B &0.201 &2101 &32.22(0.35)$^{c}$ &55782.5(10)$^{a}$ &Apr. 21 &Aug. 20.12 &53 &6 &\citet{cbet-11fd} \\
SN 2012A & NGC 3239 &0.088 &753 &29.93(0.37)$^{d}$ &55929.4(3) & Dec. 29 &Jan. 7.38 &62 &3 &\citet{cbet-12A} \\
SN 2012aw & NGC 3351 &0.076 &778 &30.01(0.09)$^{g}$ &56002.1(1)$^{f}$&Mar. 15.27 &Mar. 16.9 &284 &4 &\citet{cbet-12aw} \\
SN 2012ck & Anon. &0.260 &12520 &36.30(0.05) &56064.5(2) &May 15.50 &May 19.50 &125 &4 &\citet{cbet-12ck} \\
SN 2012ec & NGC 1084 &0.073 &1407 &31.20(0.40)$^{c}$ &56142.5(9)$^{f}$ &$\cdots$ &Aug 11.04 &77 &8$^{h}$ &\citet{cbet-12ec} \\
SN 2013ab & NGC 5669 &0.075 &1368 &31.40(0.53)$^{c}$ &56339.5(1) &Feb. 15.53 &Feb. 17.54 &165 &14 &\citet{cbet-13ab} \\
SN 2013am & NGC 3623 &0.068 &807 &30.54(0.40)$^{c}$ &56371.5(1.5)$^{f}$ &Mar. 20.20 &Mar 21.64 &86 &2 &\citet{cbet-13am} \\
SN 2013bu & NGC 7331 &0.250 &816 &30.79(0.08)$^{g}$ &56399.3(4.5) &Apr. 12.8 &Apr. 21.76 &64 &1$^{h}$ &\citet{cbet-13bu} \\
SN 2013ej$^{e}$ & NGC 628 &0.191 &657 &29.93(0.40)$^{c}$ &56496.9(1)$^{f}$ &Jul 14.42 &Jul. 25.45 &449 &8 &\citet{cbet-13ej} \\
SN 2013fp & IC 421 &0.663 &3548 &33.57(0.17) &56546.9(7.5) &Sep. 4.4 &Sep. 19.4 &55 &1 &\citet{cbet-13fp} \\
SN 2013ft & NGC 774 &0.148 &2907 &33.66(0.29)$^{d}$ &56546.8(1) &Sep. 11.29 &Sep. 13.29 &135 &1$^{h}$ &\citet{cbet-13ft} \\
SN 2013gd & MCG-01-10-39 &0.374 &4021 &33.75(0.16) &56603.3(2) &Nov. 5.3 &Nov. 9.35 &113 &2 &\citet{cbet-13gd} \\
SN 2014G & NGC 3448 &0.003 &1350 &31.94(0.80)$^{c}$ &56669.5(2) &Jan. 10.85 &Jan 14.32 &150 &5 &\citet{cbet-14G} \\
SN 2014ce & NGC 7673 &0.119 &3408 &33.22(0.20) &56877.5(1) &Aug. 8 &Aug. 9.52 &62 &1 &\citet{cbet-14ce} \\
SN 2014cn & NGC 4134 &0.05 &3826 &34.02(0.15) &56767.2(4) &Apr. 16 &Apr. 24.38 &113 &0 &\citet{cbet-14cn} \\
SN 2014cy & NGC 7742 &0.049 &1663 &31.73(0.80)$^{c}$ &56899.5(1)$^{f}$ &Aug 29.3 &Aug 31.0 &99 &8 &\citet{cbet-14cy} \\
SN 2014dq & ESO 467-G51 &0.051 &1808 &31.35(0.28)$^{c}$ &56945.6(3) &Oct. 13.09 &Oct. 19.09 &97 &4$^{h}$ &\citet{cbet-14dq} \\
SN 2015C & IC 4221 &0.223 &2889 &33.32(0.20) &57003.0(19)$^{a}$ &$\cdots$ &Jan. 7.60 &46 &4 &\citet{cbet-15C} \\
SN 2015O & PGC 1426131 &0.404 &16788 &36.98(0.04) &57194.7(1) &Jun. 21 &Jun. 22.38 &68 &2 &\citet{cbet-15O} \\
SN 2015V & UGC 11000 &0.105 &1369 &31.42(0.73)$^{c}$ &57112.5(4) &Mar. 27 &Apr. 4.52 &287 &14 &\citet{cbet-15V} \\
SN 2015W & UGC 3617 &0.380 &3984 &33.88(0.15) &57020.5(16) &14 Dec. 14 &Jan. 12.27 &62 &2 &\citet{cbet-15W} \\
SN 2015X & UGC 3777 &0.162 &3213 &33.42(0.19) &57074.1(2) & Feb. 19 &Feb. 23.22 &109 &1$^{h}$ &\citet{cbet-15X} \\
SN 2015be & NGC 1843 &0.402 &2603 &32.63(0.17)$^{c}$ &57360.2(2) &Dec. 2 & Dec. 6.39 &19 &6 &\citet{cbet-15be} \\
SN 2016X & UGC 08041 &0.061 &1321 &30.91(0.43)$^{c}$ &57406.4(1.0) &Jan 18.35 &Jan 20.58 &203 &10 &\citet{cbet-16X} \\
SN 2016adg & UGC 3376 &0.875 &3945 &33.81(0.16) &57420.2 (5.0) &Jan 18.35 &Jan 20.58 &95 &6 &\citet{cbet-16adg} \\
SN 2016cok & M66 &0.091 &727 &30.13(0.08)$^{g}$ &57534.3(2) &May 24.32 &May 28.29 &34 &1 &\citet{cbet-16cok} \\
SN 2016cyx & UGC 01814 &0.401 &4104 &33.72(0.17) &57569.6(6) &Jun. 24.60 &Jul. 6.59 &41 &1$^{h}$ &\citet{cbet-16cyx} \\
SN 2016fqr & NGC 1122 &0.242 &3599 &33.46(0.18) &57632.0(1.5) &Aug. 30.49 &Sep. 2.52 &49 &7 &\citet{cbet-16fqr} \\
SN 2017faf & Annon. &0.187 &8845 &35.54(0.08) &57930.5(2) &Jun 24.46 &Jun 28.43 &260 &6 &\citet{cbet-17faf} \\
SN 2017hta & UGCA 81 &2.867 &1338 &31.59(0.40)$^{c}$ &58054.8(5) &Oct. 24.45 &Nov. 2.31 &28 &1 &\citet{cbet-17hta} \\
SN 2017iit & UGC 3232 &1.615 &5006 &34.29(0.13) &58074.5(1) &Nov. 16.50 &Nov. 18.50 &33 &1 &\citet{cbet-17iit} \\
SN 2017jbj & NGC 259 &0.124 &4045 &33.64(0.17) &58104.1(5)$^{a}$ &$\cdots$ &Dec. 20.47 &20 &1$^{h}$ &\citet{cbet-17jbj} \\
SN 2018hde & CGCG 230-008 &0.323 &10240 &35.84(0.07) &58397.2(1) &Oct. 5.34 &Oct. 7.17 &64 &1 &\citet{cbet-18hde}\\
\hline
\label{SN_sample}
\end{longtable}
\vspace{-0.60cm}
\noindent
\hspace{-0.90cm}
\begin{minipage}[b]{19.5cm}
\scriptsize
Notes: The relevant information for all SNe~II from the Berkeley sample is displayed. The first column gives the SN name, followed by (Column 2) the name of its host galaxy and (Column 3) its reddening due to dust in our Milky Way Galaxy \citep{schlafly11}. We then (Column 4) list the host-galaxy recession velocity taken from NED and (Column 5) the distance modulus. The explosion epoch and its uncertainty are given in Column 6. Columns 7 and 8 respectively give the UT dates of the last nondetection and the discovery. Column 9 presents the number of photometric points (including $BVRI$ bands), while Column 10 gives the number of spectra. Finally, Column 11 lists the discovery reference.\\
$^{a}$ Explosion date determined using SNID \citep{blondin07}.\\
$^{b}$ Spectra taken from the literature (CSP-I; \citealt{gutierrez17b}).\\
$^{c}$ From NED using Tully-Fisher measurements. Uncertainties are the standard deviation of the mean.\\
$^{d}$ From SN measurements: NGC 918 \citep{maguire12}, SN~2012A \citep{nugent06,poznanski09,rodriguez14}, and SN~2013ft \citep{dejaeger17a}.\\
$^{e}$ LOSS data already used/published: SN~2009kr \citep{eliasrosa10}, SN~2010id \citep{galyam10}, and SN~2013ej \citep{dhungana16}.\\
$^{f}$ Information found in the literature: PTF10vld \citep{galyam10}, PTF12bvh \citep{atel-12aw}, SN~2012ec \citep{barbarino15}, SN~2013am \citep{tomasella18}, SN~2013ej \citep{dhungana16}, and SN~2014cy \citep{valenti16}.\\
$^{g}$ From Cepheid measurements: NGC 3351 \citep{graham97}, NGC 7331 \citep{kanbur03}, and M66 \citep{kanbur03}.\\
$^{h}$ Spectra taken from the literature (\url{https://wiserep.weizmann.ac.il/}).\\
\end{minipage}

\begin{figure*}
\includegraphics[width=1.0\textwidth]{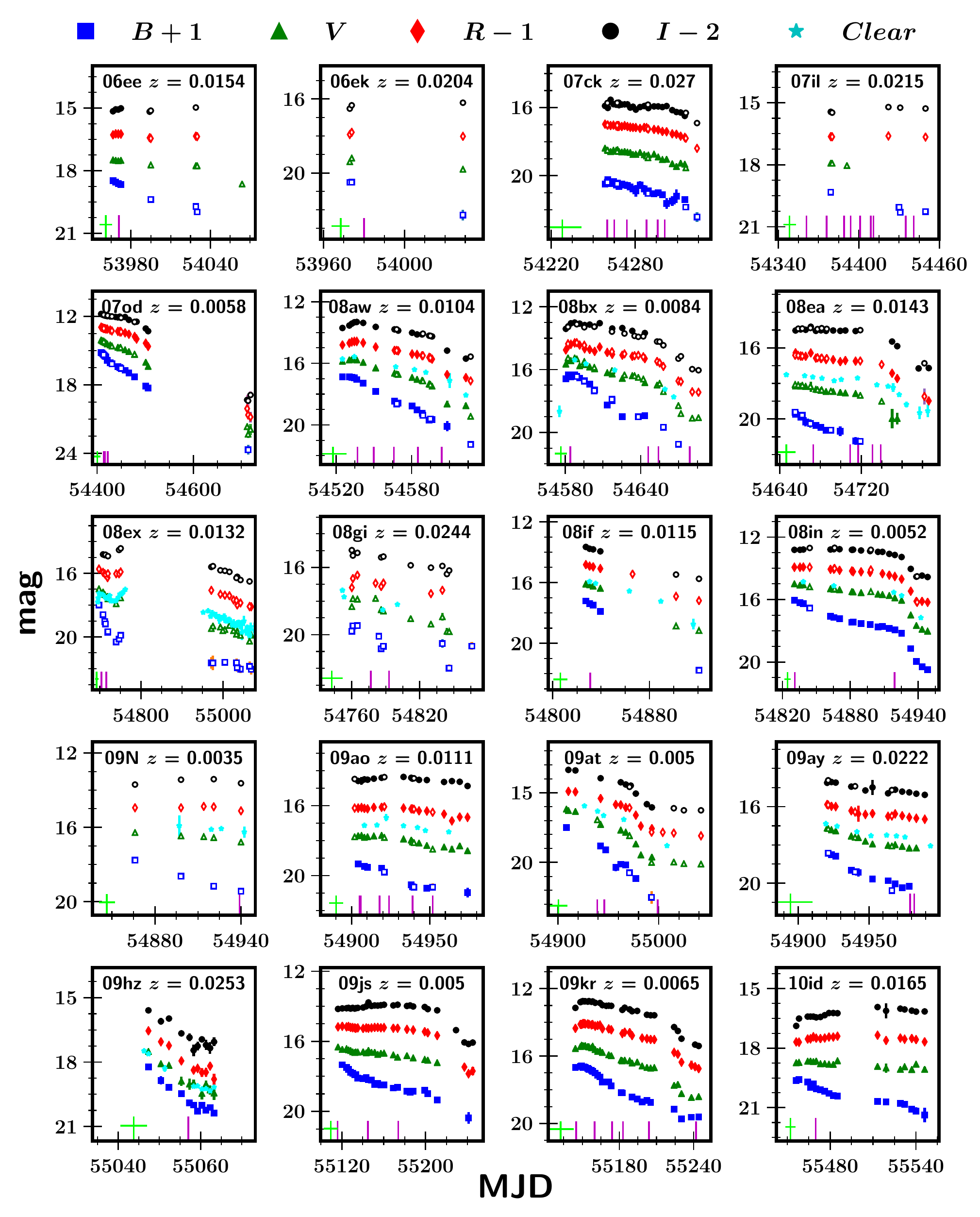}\\

\caption{SN~II observed light curves corrected for MW extinction. Blue squares are magnitudes in $B$, green triangles are $V$, red diamonds are $R$, and black circles are $I$. The abscissa is the Modified Julian Date (MJD). In each panel, the IAU name and the redshift are given in the upper right. Full symbols are KAIT data while empty symbols are Nickel data. The vertical magenta lines indicate the epochs of optical spectroscopy while the vertical green line represents the explosion date and its associated uncertainty.}
\label{fig:LC_KAIT}
\end{figure*}

\begin{figure*}
\addtocounter{figure}{-1}

\includegraphics[width=1.0\textwidth]{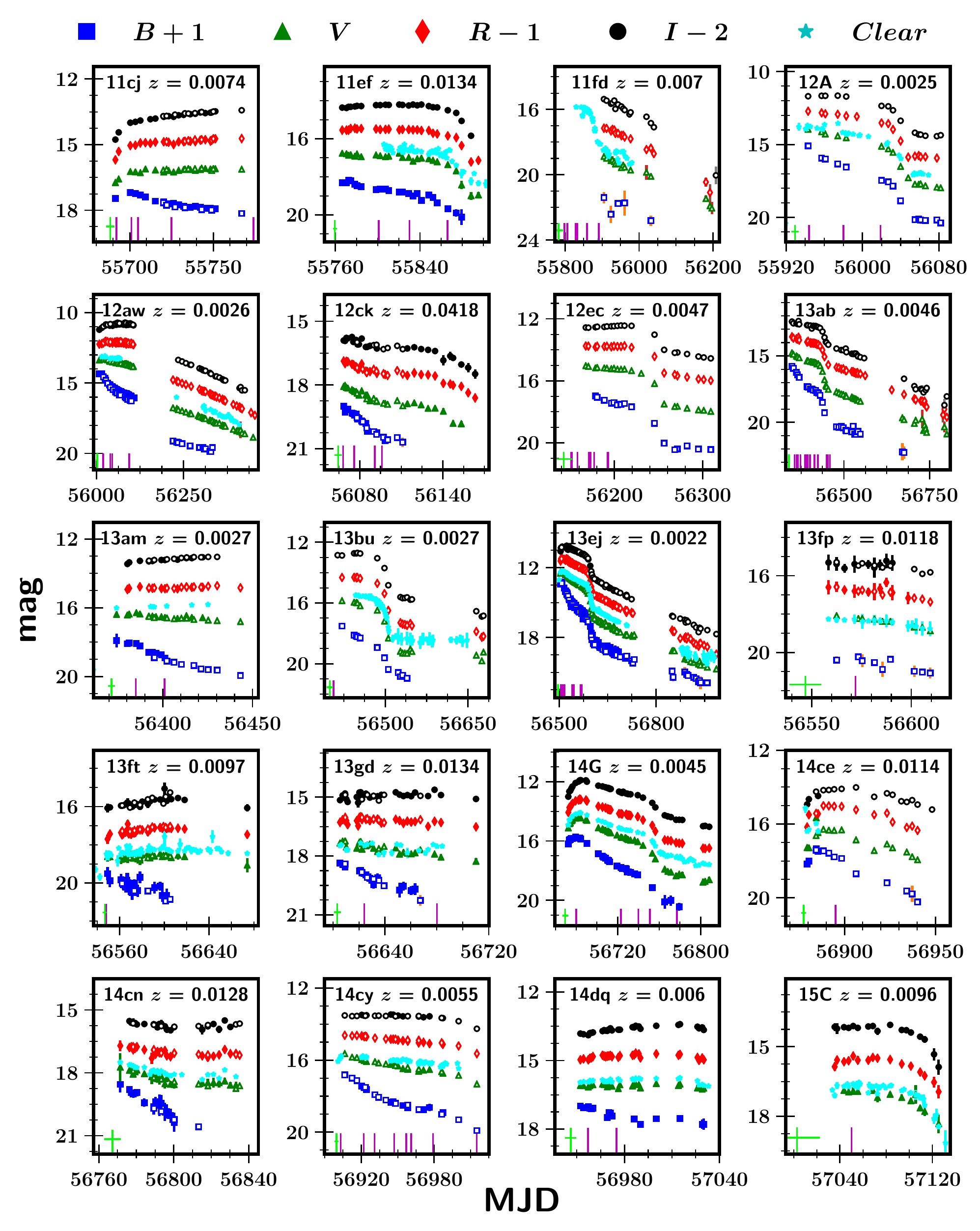}\\

\caption{SN~II observed light curves corrected for MW extinction. Blue squares are magnitudes in $B$, green triangles are $V$, red diamonds are $R$, and black circles are $I$. The abscissa is the Modified Julian Date (MJD). In each panel, the IAU name and the redshift are given in the upper right. Full symbols are KAIT data while empty symbols are Nickel data. The vertical magenta lines indicate the epochs of optical spectroscopy while the vertical green line represents the explosion date and its associated uncertainty.}
\label{fig:LC_KAIT}
\end{figure*}

\begin{figure*}
\addtocounter{figure}{-1}

\includegraphics[width=1.0\textwidth]{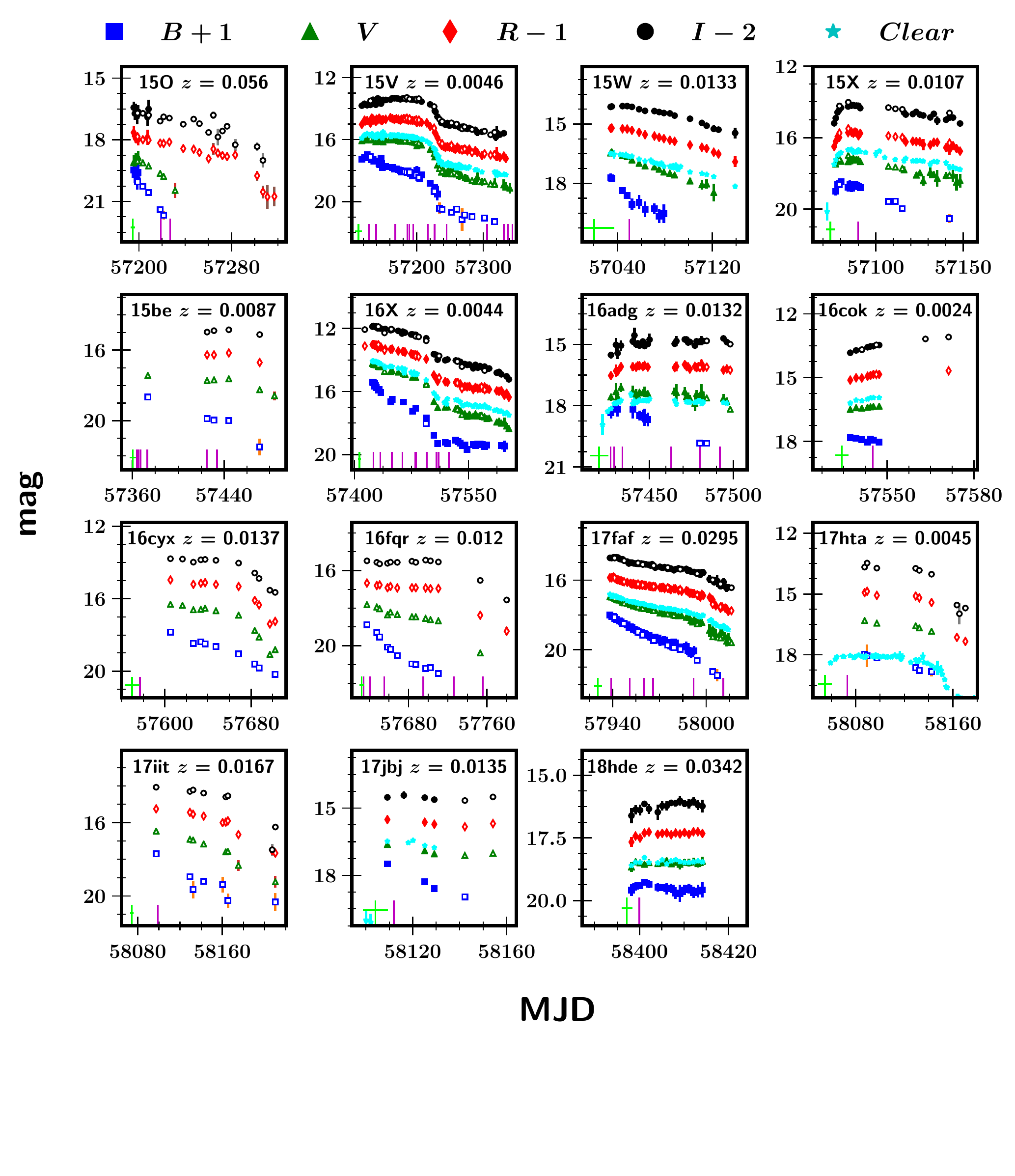}\\

\caption{SN~II observed light curves corrected for MW extinction. Blue squares are magnitudes in $B$, green triangles are $V$, red diamonds are $R$, and black circles are $I$. The abscissa is the Modified Julian Date (MJD). In each panel, the IAU name and the redshift are given in the upper right. Full symbols are KAIT data while empty symbols are Nickel data. The vertical magenta lines indicate the epochs of optical spectroscopy while the vertical green line represents the explosion date and its associated uncertainty.}
\label{fig:LC_KAIT}
\end{figure*}

\begin{figure*}
\includegraphics[width=1.0\textwidth]{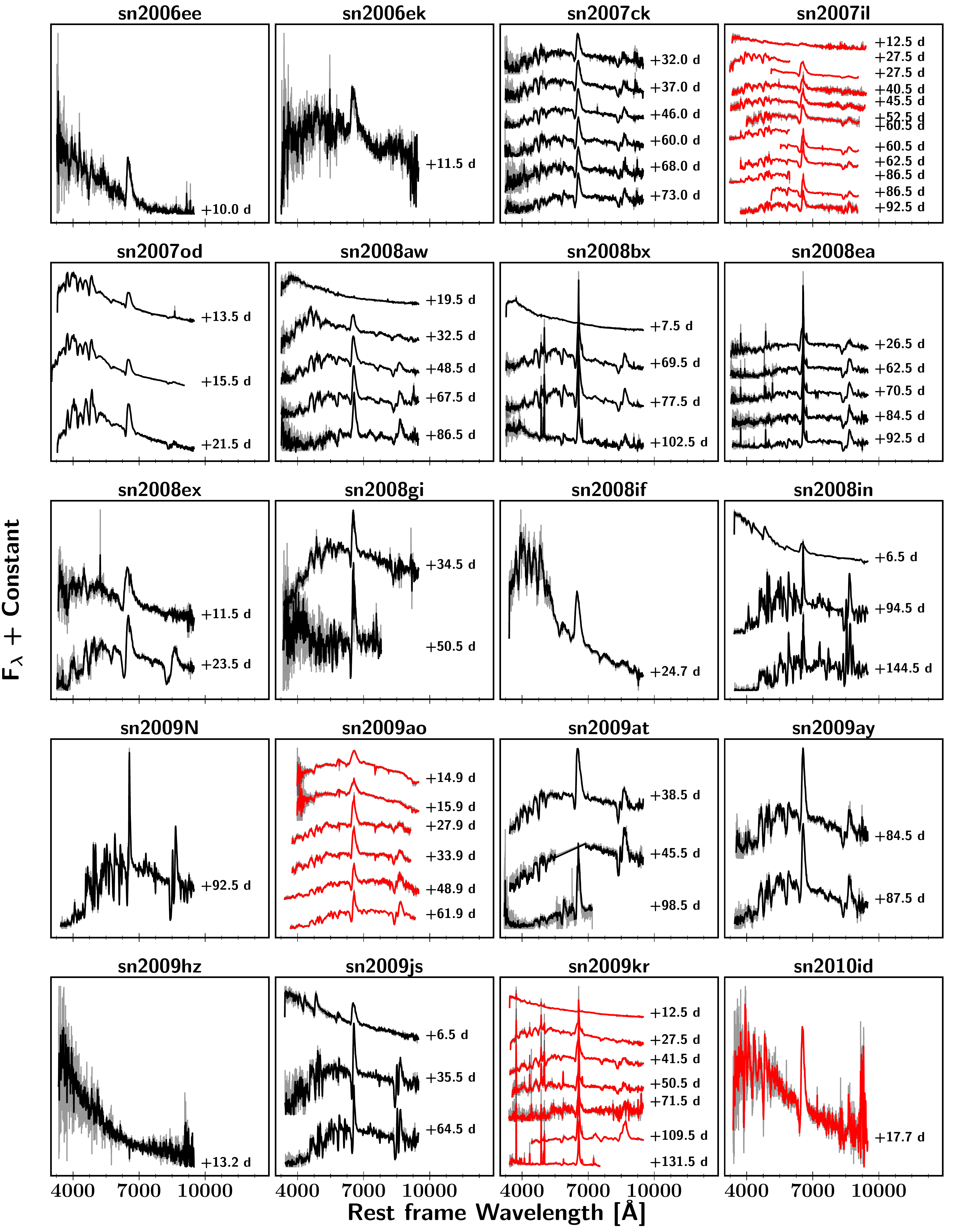}

\caption{Spectral sequence for each SN. The spectra are shown in the rest frame, and the date listed for each SN is the number of days since the explosion (rest frame). The redshift of each SN is also labelled. The original spectra are shown in grey while in black the spectra are binned (10~\AA). We represent in red the spectra that are already available the literature.}
\label{fig:spec_KAIT}
\end{figure*}

\begin{figure*}
\addtocounter{figure}{-1}
\includegraphics[width=1.0\textwidth]{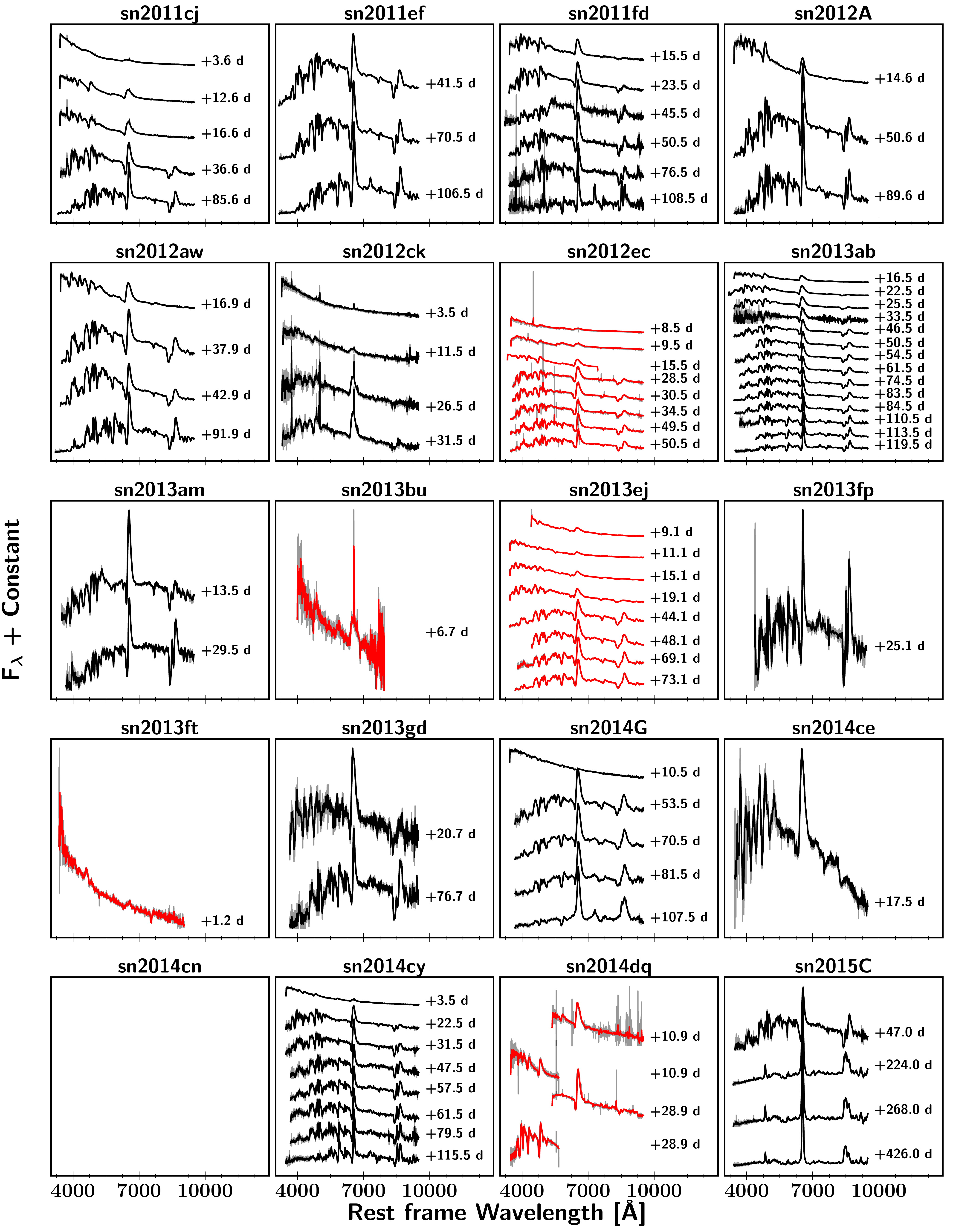}

\caption{Spectral sequence for each SN. The spectra are shown in the rest frame, and the date listed for each SN is the number of days since the explosion (rest frame). The redshift of each SN is also labelled. The original spectra are shown in grey while in black the spectra are binned (10~\AA). We represent in red the spectra that are already available the literature.}
\label{fig:spec_KAIT}
\end{figure*}

\begin{figure*}
\addtocounter{figure}{-1}

\includegraphics[width=1.0\textwidth]{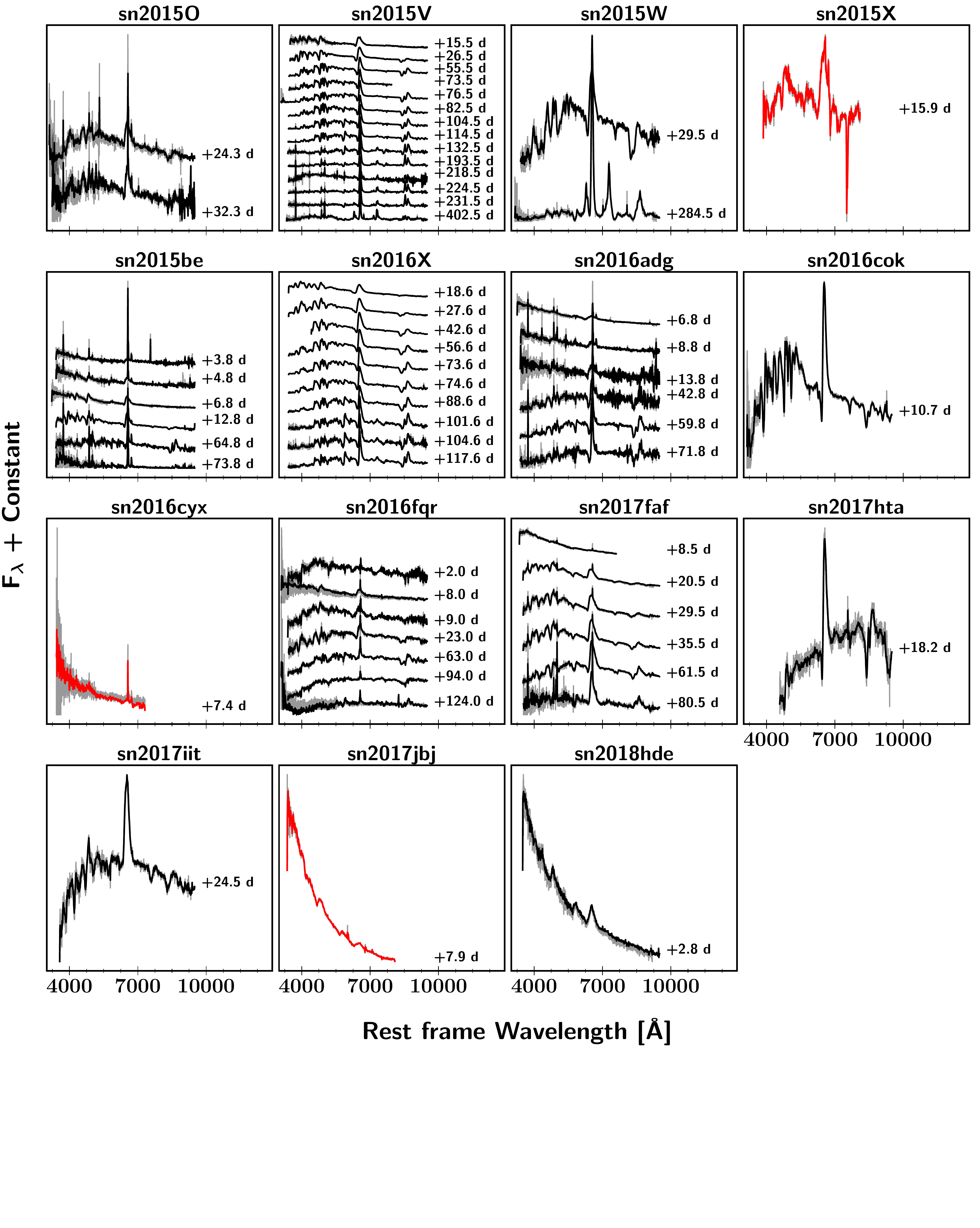}

\caption{Spectra sequence for each SNe. The spectra are shown in the rest frame, and the date
listed for each SN is the number of days since the explosion (rest frame). The redshift of each SN is also labelled. The original spectra are shown in grey while in black the spectra were binned (10\AA). We represent in red the spectra that are already available the literature.}
\label{fig:spec_KAIT}
\end{figure*}

\bsp	
\label{lastpage}
\end{document}